\documentclass[12pt]{article}
\usepackage[T1]{fontenc}
\usepackage{geometry}
\usepackage{natbib}
\usepackage{float}
\usepackage{tcolorbox}
\usepackage{arydshln}
\usepackage{bm}
\usepackage{setspace}
\usepackage{tikz}
\usepackage{bbm}
\usepackage{adjustbox}
\usepackage{algorithm}
\usepackage{graphics}
\usepackage{xcolor}
\usepackage{arydshln}
\usepackage{stackengine}
\stackMath
\newcommand\tsup[2][2]{%
	\def\useanchorwidth{T}%
	\ifnum#1>1%
	\stackon[-.5pt]{\tsup[\numexpr#1-1\relax]{#2}}{\scriptscriptstyle\sim}%
	\else%
	\stackon[.5pt]{#2}{\scriptscriptstyle\sim}%
	\fi%
}

\usepackage{amsmath}
\usepackage{amsthm}
\usepackage{amssymb}
\usepackage{booktabs,caption}
\usepackage{threeparttable}
\usepackage{graphicx}
\usepackage{color}
\providecommand{\keywords}[1]{{\textit{Keywords---}} #1}
\providecommand{\JELclassification}[1]{{\textit{JEL---}} #1}
\usepackage{bbm}
\makeatletter
\theoremstyle{plain}
\newtheorem{thm}{\protect\theoremname}
  \theoremstyle{plain}
  \newtheorem{assumption}{\protect\assumptionname}[section]
  \theoremstyle{plain}
  \newtheorem{prop}{\protect\propositionname}[section]
  \theoremstyle{plain}
  \newtheorem{lem}{\protect\lemmaname}[section]
  \theoremstyle{plain}
  
  \theoremstyle{plain}

\newtheorem{definition}{Definition}[section]
\makeatother

\usepackage[english]{babel}
  \providecommand{\assumptionname}{Assumption}
  \providecommand{\lemmaname}{Lemma}
  \providecommand{\propositionname}{Proposition}
\providecommand{\theoremname}{Theorem}
\usepackage{thmtools}
\renewcommand\thmcontinues[1]{Continued}
\usepackage{authblk}
\numberwithin{equation}{section}

\usepackage[hidelinks,colorlinks=true, citecolor=blue]{hyperref} 
\begin{document}
	\title{Robust Bayesian Method for Refutable Models}
	\author[1]{Moyu Liao \thanks{Email: moyu.liao@sydney.edu.au. First draft: January, 2024.}}
	\affil[1]{The University of Sydney}
	\maketitle

	\begin{abstract}
		\begin{spacing}{1.2}

{		We propose a robust Bayesian method for economic models that can be rejected by some data distributions. The econometrician starts with a refutable structural assumption which can be written as the intersection of several assumptions. To avoid the assumption refutable, the econometrician first takes a stance on which assumption $j$ will be relaxed and considers a function $m_j$ that measures the deviation from the assumption $j$. She then specifies a set of prior beliefs $\Pi_s$ whose elements share the same marginal distribution $\pi_{m_j}$ which measures the likelihood of deviations from assumption $j$. Compared to the standard Bayesian method that specifies a single prior, the robust Bayesian method allows the econometrician to take a stance only on the likeliness of violation of assumption $j$ while leaving other features of the model unspecified.  We show that many frequentist approaches to relax refutable assumptions are equivalent to particular choices of robust Bayesian prior sets, and thus we give a Bayesian interpretation to the frequentist methods.  We use the local average treatment effect ($LATE$) in the potential outcome framework as the leading illustrating example. 
}

		\end{spacing}
	

	\JELclassification{ C12, C13, C18, C51, C52}

	\end{abstract}

	\pagebreak
	
	\section{Introduction}
		Model assumptions are often crucial for identification in many econometric models. Often, applied researchers use convenient econometric assumptions to derive an informative identified set of parameters of interest. However, assumptions may be rejected by some data distribution, which leads to an empty identified set. When such a problem arise, we call the assumption refutable. To facilitate the discussion, suppose that the original refutable assumption $A$ can be written as the intersection of several weaker assumptions, i.e., $A=\cap_{l} A_l$, and the econometrician knows that $A_j$ will lead to the refutation issue.

		When an economic assumption is refuted, there are several ways to salvage the refutable assumptions.  Frequentist has two major ways to deal with a refutable assumption: The first approach is to completely give up $A_j$ that can lead to data rejection and then identify the parameter of interest under the weaker assumption. However, this approach often gives up the original economic theory behind the assumption. As a result, we may get a wide identified set that is not informative of the parameter of interest. Another approach is to minimally relax $A_j$ until the relaxed assumption cannot be rejected and then identify the parameter of interest under the minimally relaxed assumption. This approach is similar to the sensitivity analysis. However, it is hard to justify the rationality of the minimally relaxed assumption and the interpretation of the consequent indentified set becomes vague. 
		 
		In contrast, the Bayesian approach of dealing with refutable assumption can deliver better interpretation of the relaxed assumption. The Bayesian econometrician first specifies a prior belief over all econometric structures. To incorporate the refutable assumption as a reasonable economic assumption, the econometrician may specify a prior belief with a higher prior probability on the original assumption.  However, specifying a unique prior belief requires the econometrician to take a prior stance not only on the likelihood of violation of $A_j$, but also on all other aspects of the model.  As a result, specifying a unique prior belief may overstate the econometrician's belief over all possible econometric structures. 
	
		This paper proposes using a robust Bayesian method to deal with refutable assumptions. Compared to the standard Bayesian method, our robust Bayesian method allows the econometrician to consider a set of priors that has a common marginal distribution.  This common marginal distribution measures the econometrician's belief of how likely $A_j$ is violated. As a result, the econometrician can choose to only take a stance on devation from $A_j$ while remain agnostic about other aspects of the econometric model. There are several appealing features of our robust Bayesian method. First, it can be shown that under proper choices of the common marginal distribution, the robust Bayesian method is equivalent to certain frequentist approaches to salvage the refutable model. More specifically, the robust Bayesian method estimated identified set will converge to the frequentist identified set.\footnote{We do not seek the equivalence result for inference because many econometric models are nonparametric.} As a result, we can interpret the frequentist approaches to the refutable models as a choice of the robust Bayesian prior set. Second, unlike the standard Bayesian method which often generates a point estimate from the posterior, the robust Bayesian method can generate a posterior mean set which is similar to the partial identification method. Third, in many nonparametric models, the common marginal constraint allows us to find computationally feasible ways to compute the posterior mean set. 
	
		To preview the robust Bayesian framework, we first formalize some essential notations in the previous discussion: Recall that the original refutable assumption $A$ as the countable intersection of assumptions $A=\cap_{l=1}^\infty A_l$, where each assumption $A_l$ is a collection of econometric structures $s$. Each econometric structure $s$ specifies an economic mechanism and predicts a unique data distribution $F$. The Bayesian econometrician takes a stance on which assumption $A_j$ is likely to be violated and only relaxes $A_j$. For the assumption $A_j$ that will be relaxed, the econometrician choose a distance function $m_j(s')$ that measures the distance of any econometric structure $s'$ to $A_j$.  Since each econometric structure $s$ predicts a unique data distribution,  each prior belief $\pi_s$ over structures induce a joint distribution of $(s,F,m_j(s))$. To avoid over-specification of her belief, the econometrician considers a particular set of beliefs $\Pi_s$ whose elements are supported on $\cap_{l\ne j} A_l$ with the constraint that all $\pi_s\in \Pi_s$ induce the same conditional marginal conditional marginal distribution $\pi^*_{m_j|F}$. By choosing to further condition $\pi^*_{m_j|F}$ on $F$, we can allow more flexible characterization of the prior beliefs and we can show some equivalence results to the frequentist-based approaches.
	
		After observing realized data $\bm{X}_n$, the econometrician updates each belief in the prior set and collect the posteriors as  $\Pi_{s|\bm{X}_n}$. We follow \cite{giacomini2021robust} to study two statistics from the posterior set. The first one is posterior-mean bound. For each posterior belief in the posterior set, we can derive the posterior mean of the parameter of interest. By maximizing and minimizing the posterior mean of the parameter of interest, we can derive a bound. It can be shown that by choosing $\pi^*_{m_j|F}$ to be a proper degenerate marginal distribution, the posterior mean bound is the convex hull of the identified set under the frequentists' minimally relaxed assumption approach. Characterizing the maximum and minimum can be hard because we are maximizing over a set of posterior distributions. We then show that we can transform the optimization problem to an optimization over all structures with a fixed amount of deviation, i.e. $m_j(s)=m$, and then integrate a corresponding parameter quantity with respect to the joint posterior belief over $(m_j(s),F)$.  The second quantity is a confidence interval that covers the parameter of interest with $1-\alpha$ probability under all posterior beliefs. The calculation of the confidence interval is similar to that of the posterior-mean bound, which uses the quantiles of the parameter of interest under the  joint posterior belief on $(m_j(s),F)$.
	
		For the main application,  we look at the local average treatment effect ($LATE$) in the potential outcome framework. \cite{kitagawa2015} and \cite{mourifie2017testing} show the `No Defiers' assumption and the independent IV assumption are jointly refutable. We  proceed to consider a prior belief set with marginal density constraints on the measure of defiers. Since each econometric structure in the $LATE$ framework contains a distribution of potential outcomes which is infinite dimensional, the optimization to find the posterior mean bound of $LATE$ is an infinite dimensional optimization problem. However, we use several tricks to convert the infinitely dimensional optimization problem to a combination of a feasible finite dimensional optimization problem and a integratoin problem. We apply the robust Bayesian method to study the return of college education \citep{card1993using} and compare our result with that of frequentists' approach. In addition to the main application, we also look at the applications to monotone IV model, the intersection bounds model and the Logit discrete choice model. 
	
		The framework in this paper is mostly related to the robust Bayesian method in \cite{giacomini2021robust} but we add discussion to their paper in several ways. First, \cite{giacomini2021robust} show that the posterior mean set is equivalent to the frequentist-based partially identified set and the variation of the prior set in their framework does not matter asymptotically. In particular, when the model is parametric, they establish a Berstein-von Mises result for the robust Bayesian method.\footnote{In principle, if only convergence is required, their method can accomodate nonparametric models.}  In contrast, we use the robust Bayesian method as a modeling choice to deal with refutable assumptions. More precisely, the choice of the marginal distribution of $m_j$ matters for the final posterior mean bound and the confidence set.\footnote{From this perspective, our method has a Bayesian interpretation because the econometrician has to take a stance on which assumption can fail and how severely the assumption might be violated.} We show some equivalence to some frequentist-based approaches, but we only aim to show that these frequentist-based approaches have a Bayesian interpretation. Second, it is computationally hard to accomodate nonparametric models in \cite{giacomini2021robust} because of the specific choice of the prior set. In contrast, by further imposing the unique marginal constraint $\pi_{m_j|F}^*$,  we can simplify the computation in many econometric settings. However, we do not establish the nonparametric Berstein-von Mises result.

	\paragraph{Related Literature} Our paper contributes to the literature on salvaging refutable or misspecified models. \citet{masten2018} propose an ex-post way to salvage refutable models by considering a minimal relaxation method. \citet{bonhomme2018minimizing} consider a local sensitivity analysis of a potentially misspecified model.  \citet{christensen2023counterfactual} consider a sensitivity analysis of the counterfactuals to parametric model assumptions.
	
	We also contribute to the use of the robust Bayesian method. \cite{giacomini2021robust} show the partial-identification version of Bernstein-von-Mises equivalence theorem for parametric model. The robust Bayesian method is also applied to study the uncertain identification in SVARs model \cite{giacomini2022robust}. In statistics, the robust Bayesian method has been studied by \cite{deroberts1981bayesian}, \cite{berger1986robust}, \cite{wasserman1989robust}, \cite{wasserman1992computing}. 
	
	Our application to the $LATE$ framework contributes to the study of $LATE$ under the `No Defiers' condition. { Since \citet{kitagawa2015} proves the sharp testable implication of \citet{IA1994}, literature relaxes the `No Defiers' condition. \citet{chaisemartin2018Defier} discusses the economic meaning of the conventional $LATE^{Wald}$ when there are defiers. He shows that the $LATE^{Wald}$ identifies the net average treatment effect of a subgroup of compliers after deducting the average treatment effect of defiers.}

	
	The rest of the paper is organized as follows. Section \ref{section: $LATE$application} uses the potential outcome model \citep{IA1994} to introduce the robust Bayesian method. We use this section as a concrete example to illustrate the robust Bayesian method procedures and also the possible difficulties.  Section \ref{section: General Theory} describes a theory of robust Bayesian method. We also show the general frequentist-equivalence result in this section.  Section \ref{sec: Additional Applications} considers further applications. Section \ref{section: Conclusion} concludes. All proofs are collected in Appendices. 
		
	\section{A Leading Example of the Potential Outcome Model}\label{section: $LATE$application}

We start with an example of a binary treatment and a binary instrument  \citep{IA1994}. This application illustrates the construction of the robust Bayesian prior set and the connnection to frequentists' approach. Notations are set up to match the general theory in the later sections. 


\subsection{The Potential Outcome Framework}
An econometrician observes an outcome variable $Y_i$, a treatment decision $D_i$, and a binary instrument $Z_i$. The observed variables $(Y_i,D_i)$ are generated through the following potential outcome framework:
\begin{align} \label{eq: potential outcome}
\begin{split}
Y_i&= Y_i(1)D_i+Y_i(0)(1-D_i),\\
D_i&= D_i(1)Z_i +D_i(0)(1-Z_i),
\end{split}
\end{align}
where $D_i(1),D_i(0)$ are potential treatment decisions, $Y_i(1),Y_i(0)$ are the potential outcomes and $Z_i$ is a binary instrument. We implicitly impose the exclusion restriction in (\ref{eq: potential outcome}).

Variables in (\ref{eq: potential outcome}) can be classified into two types. The variables $\epsilon_i=(D_i(1), D_i(0), Y_i(0), Y_i(1), Z_i)$  reflect the fundamental heterogeneity of the economic agent $i$ and we call them the underlying variables. The variables $X_i=(Y_i, D_i, Z_i)$ are observed variables. Let $\mathcal{Y}$ be the space of $Y_i$ and let $\mathcal{B}$ be the Borel-sigma algebra on $\mathcal{Y}$. We consider two distribution spaces: The space of distributions of $X_i$ is 
\begin{equation}\label{eq: appli, observation space}
\mathcal{F}=\{F_X(y,d,z):\,\, y\in\mathcal{Y},\, d,z\in\times\{0,1\}\},
\end{equation}
and space of distributions of underlying variables is
\begin{equation}\label{eq: appli, primitive space}
\begin{split}
\mathcal{G}=\big\{&G(\epsilon) : \epsilon\in\{0,1\}^2\times\mathcal{Y}^2\times \{0,1\}  \big\}.
\end{split}
\end{equation}
Starting from a $G^s(\epsilon)$, the potential outcome equation (\ref{eq: potential outcome}) defines a unique distribution of observable via the map $M:\mathcal{G}\rightarrow \mathcal{F}$ such that: 
\begin{equation}
\begin{split}
M(G^s)=\big\{F\in \mathcal{F}:\, &\forall B\in \mathcal{B}, \quad d,z\in\{0,1\},\\
 Pr_{F}(Y_i&\in B,D_i=d,Z_i=z)= Pr_{G^s}(Y_i(d)\in B,D_i(z)=d,Z_i=z)
\big\}.\label{eq: appli, M^s}
\end{split}
\end{equation}
In other words, $F$ is the push-forward distribution of $G^s$ under the mapping (\ref{eq: potential outcome}). We call a pair $(M,G^s)$ an econometric structure.
In contrast to the $G^s$ that describes the individual heterogeneity, the mapping $M$ describes the economic machinsm to generate observed varaibles. While $M$ is fixed for the potential outcome model, it can vary in the later general theory section.\footnote{Think of an OLS regression $Y_i=\beta_0+\beta_1 X_i+\epsilon_i$, the mapping $M$ depends on the parameter $(\beta_0,\beta_1)$.} We consider a paradigm for analysis of the model:
\begin{equation} \label{eq: appli, potential outcome model space}
	\mathcal{S}=\left\{s|\,\, G^s\in \mathcal{G},\,\, M \,\,\text{satisfies} \,\, (\ref{eq: appli, M^s}) \right\}.
\end{equation}

Empirical researchers often use the Imbens-Angrist Monotonicity assumption (IA-M) where the exogeneity and monotonicity of the instrument $Z_i$ are assumed. We formalize the IA-M assumption (denoted by $A$ ):
\begin{equation} \label{eq: IA instrument assumption }
\begin{split}
A&=A^{ND}\cap A^{IV}\\
A^{ND}&=\{s: D_i(1)\ge D_i(0), G^s\text{-a.s.}\}\\
A^{IV}&=\{s: G^s \text{ satisfies } Z_i\perp (Y_i(1),Y_i(0),D_i(1),D_i(0))\},
\end{split}
\end{equation}
where $A^{ND}$ is the `No Defiers' assumption and $A^{IV}$ is the independent IV assumption.  The main parameter of interest is the local average treatment effect for compliers:
\begin{equation}\label{eq: appli, formular}
	LATE(G^s)\equiv E[Y_i(1)-Y_i(0)|D_i(1)=1,D_i(0)=0].
\end{equation}
The IA-M assumption is preferred by applied econometricians as the economic intuition behind it is clear: Defiers are abnormal and assumed away from the model. If the IA-M assumption holds,  we can identify the $LATE$ as the Wald ratio \citep{IA1994}:
\[
LATE^{Wald}(F)=\frac{E_F[Y_i|Z_i=1]-E_F[Y_i|Z_i=0]}{E_F[D_i|Z_i=1]-E_F[D_i|Z_i=0]}.
\]

\subsubsection{The sharp testable implication of $A$}
The IA-M assumption can be rejected by some data distributions. We summarize the sharp testable implications in \cite{kitagawa2015} who define the following two quantities for all $B\in \mathcal{B}$ and $d\in \{0,1\}$:
\begin{equation}\label{eq: potential outcome, link between F and G}
	\begin{split}
		P_F(B,d)\equiv Pr_F(Y_i\in B,D_i=d|Z_i=1), &\\
		Q_F(B,d)\equiv Pr_F(Y_i\in B,D_i=d|Z_i=0).&\\
	\end{split}
\end{equation}
\begin{lem}\label{lem: testable implication of IA assumption}
	Let $P_F(\cdot,d)$ and $Q_F(\cdot,d)$, $d\in\{0,1\}$, be absolutely continuous with respect to some measure $\mu_F$. For any structure $s\in A$, $F=M(G^s)$, and any Borel set $B$, the $F$ must satisfy: \begin{equation} \label{eq: testable implication of LATE}
	\begin{split}
	P_F(B,1)&\ge Q_F(B,1),\\
	Q_F(B,0)&\ge P_F(B,0).
	\end{split}
	\end{equation}
	Moreover, for any $F$ satisfying (\ref{eq: testable implication of LATE}), there is an $s\in A$ such that $F=M(G^s)$. 
\end{lem}
\noindent An implication of (\ref{eq: testable implication of LATE}) is that $E_{F}[D_i|Z_i=1]\ge E_{F}[D_i|Z_i=0]$ must hold. Whenever \eqref{eq: testable implication of LATE} fails, the identified $LATE$ is an empty set.


%


\subsubsection{Frequentist approach to the refutable IA-M assumption}
There are two approaches used by frequentists to address the issues when the baseline assumption is refutable. We focus on relaxing the `No Defiers' assumption as the illustration. The first approach is to interpret the identified expression under the relaxed assumption. In the $LATE$ example, researchers give up the `No Defiers' assumption and interpret the Wald ratio $LATE^{Wald}$ as compliers' treatment effect net of defiers' treatment effect \citep{chaisemartin2018Defier}. The second approach is to take an adaptive method to relax the assumption but still identify the average treatment for the compliers \citep{dahl2023nevertoolate,liao2020estimating}. In this approach, the econometrician first uses the data to select a model\footnote{In particular, the model with a minimal amount of defiers that can rationalize the data distribution. } that permits the prescence of defiers, and then estimate the treatment effect for compliers.  

If we take the first approach and give up the `No Defiers' assumption, we also lose the interpretation of the local average treatment for compliers. By giving up the `No Defiers' assumption completely, the identified set of the $LATE$ for compilers is generally unbounded, even if our data distribution passes the test implied by Lemma \ref{lem: testable implication of IA assumption}. By doing so, the econometrician gains the maximal robustness for the presence of defiers, but she also ignores the economic intuition behind the `No Defiers' assumption. 

The adaptive approach to relax the assumption is appealing since it keeps the economic rationales of the `No Defiers' assumption and let the data tells the probability of defiers. In particular, when the data distribution passes the test implied by Lemma \ref{lem: testable implication of IA assumption}, the data-selected model will identify the same $LATE$ quantity as the Wald ratio. However, the selection criteria are arbitrary and it may be hard to justify the choice of minimal defier with a consistent statistical framework.\footnote{\cite{liao2020estimating} interprets the selection criteria as a mixture of Bayesian model selection and a subsequent frequentist estimation stage. }

\subsection{The Robust Bayesian Approach}
\subsubsection{An alternative representation of the IA-M assumption}
Since the `No Defiers' assumption can lead to testable implications, we may want to relax $A^{ND}$ while keep the $A^{IV}$ assumption. In view of the Bayesian method, we may want to put a prior belief $\pi_s$ supported on $A^{IV}$. However, such a prior $\pi_s$ can have a critical issue, i.e., the $\pi_s$-induced $\pi_F$ is not supported on the whole $\mathcal{F}$: Let $\pi_F$ be the $\pi_s$-induced belief on $\mathcal{F}$ such that for any $\mathcal{F}_1\subseteq\mathcal{F}$, $\pi_F(\mathcal{F}_1)=\int_{\mathcal{S}} \mathbbm{1}(M(G^s)\in \mathcal{F}_1)d\pi_s$, then there exists a non-trivial\footnote{We illustrate the non-trivial subset via a simple example in \cite{liao2020estimating}. Suppose $Y_i\in\{0,1\}$ is also binary, then $P_F(Y_i=1,D_i=0|Z_i=0)\ge P_F(Y_i=1,D_i=0|Z_i=1)-P_F(D_i=1|Z_i=0)$ must hold for the observed data distribution. Now, since $Y_i$ is binary, we can characterize $F$ using an 8-d vector. It is easy to see that the above testable implication fails in a Lebesgue-positvely-measured set. } subset of $\mathcal{F}'\subseteq \mathcal{F}$ such that $\pi_F(\mathcal{F}')=0$.  This problem arises because the $A^{IV}$ still generates testable implications \citep{kitagawa2009identification}.

Instead, we consider an alternative characterization of the IA-M assumption in \cite{liao2020estimating}, under which, when we give up the $A^{ND}$, is non-refutable.
\begin{lem} \label{lem: alternative representation of IA-M assumption}
	The IA-M assumption defined in (\ref{eq: IA instrument assumption }) can be equivalently written as the intersection: $A= A^{TI}\cap A^{EM-NTAT}\cap A^{ND}$ where:
	\begin{enumerate}
		\item $A^{TI}=\left\{s\big|\, Z_i\perp \left(Y_i(0),Y_i(1)\right)|D_i(1),D_i(0) \right\}$ is the conditional type independent instrument assumption;
		\item Assumption $A^{EM-NTAT}$ is the set of structures $s$ such that: \[
		\begin{split}
			E_{G^s}[\mathbbm{1}(D_i(1)=D_i(0)=1)|Z_i=1]&=E_{G^s}[\mathbbm{1}(D_i(1)=D_i(0)=1)|Z_i=0],\\
			E_{G^s}[\mathbbm{1}(D_i(1)=D_i(0)=0)|Z_i=1]&=E_{G^s}[\mathbbm{1}(D_i(1)=D_i(0)=0)|Z_i=0].
		\end{split}\]
		This assumption says that the measure of always/never takers is independent of the instrument.
	\end{enumerate}
\end{lem}
We delegate further discussions of the above alternative repesentation to \cite{liao2020estimating}. In view of the Bayesian framework, if we instead impose a prior $\pi_s$ that is supported on $A^{TI}\cap A^{EM-NTAT}$, the induced $\pi_F$ is supported on $\mathcal{F}$.

\subsubsection{A Problem with the standard Bayesian method}
Think of an econometrician who works with the $LATE$ framework and is aware of the testable implication of the IA-M assumption. She believes that there are some economic reasons that will lead to the presence of defiers. She also believes that the presence of defiers is abnormal and is unlikely to happen. 

She decides to put a prior supported on the set $A^{TI}\cap A^{EM-NTAT}$, denoted by $\pi_{s}$. However, by doing so, she is forced to take a stance on other aspects of the distribution of potential outcomes. For example, the $\pi_{s}$ induces a distribution of the proportion of always takers:
\[ 
\pi_{AT;s}(B)\equiv \int_{\mathcal{S}} \mathbbm{1}\left(E_{G^s}[\mathbbm{1}(D_{i}(1)=D_{i}(0)=1)]\in B \right)d\pi_s.
\]
The econometrician only believes that the presence of defiers is the cause of the possible model rejection, and she is worried that taking a stance on other aspects of the distribution of potential outcomes may lead to misleading or overconfident identified quantity. Therefore, the standard Bayesian method may not suit her goal  in this context.

\subsubsection{The robust Bayesian prior set}
If a single prior can lead to over-specification of the prior belief, then the solution is to consider many prior beliefs to provide additional robustness, which is called the robust Bayesian method. We follow the robust Bayesian literature \citep{giacomini2021robust} to assume that the observed data distribution $F$ is a reduced-form parameter, and the distribution $G^s$ is the structural parameter. Two structures $s$ and $s'$ are observationally equivalent if $M(G^s)=M(G^{s'})$. For any prior belief $\pi_s$, since $\mathcal{G}$ is a polish space with respect to the total variation metric\footnote{The polishness of the space $\mathcal{G}$ is sufficient for a well-defined conditional distribution $\pi_{s|F}$. See Chapter 2 of \cite{ghosal2017fundamentals}.}, we can decompose the prior as:
\begin{equation}
\pi_{s}= \pi_{s|F} \times \pi_F,  \label{eq: decomposition of prior believes}
\end{equation}
where $\pi_F$ is a belief over the possible distributions of observed data, and $\pi_{s|F}$ is the conditional distribution of structures given observed $F$. For the belief to be consistent with the structural model, we require the posterior $\pi_{s|F}$ to be supported on the set $\{s: F=M(G^s)\}$. Following \citet{giacomini2021robust}, we call the $\pi_{F}$  the updatable part because the belief of data distribution can be updated by observing the data realization. The $\pi_{s|F}$ is called the non-updatable part because given the $F$, data observation is ancillary to the conditional belief. 

We assume that the econometrician has a unique belief marginal $\pi_F$ but can have multiple $\pi_{s|F}$ to be robust against the over-specification of the belief of structures. The uniqueness of $\pi_F$ does not harm the Bayesian modeling but helps with computation: The frequentist-Bayesian equivalence will ensure the posterior of $F$ does not depend on $\pi_F$ in the limit. Since the econometrician believes that the presence of defiers is the main reason for model rejection, she may only want to discipline the `No Defiers' aspect of the $G^s$ distribution. Given a prior $\pi_s$, let the conditional marginal distribution of amount of  defiers be:
\begin{eqnarray}
\pi_{DF;s|F}(B)&\equiv \int \mathbbm{1}\left(m^{df}(G^s)\in B \right)d\pi_{s|F}(G^s),\label{eq: marginal distribution of defiers}\\
m^{df}(G^s)&\equiv E_{G^s}[\mathbbm{1}(D_i(1)=0,D_i(0)=1)]. \label{eq: definition of measure of defiers}
\end{eqnarray}

The econometrician forms a prior belief set $\Pi_s$ that has the following representation:
\begin{equation}\label{eq: prior set}
\begin{split}
\Pi_{s}=\Bigg\{\pi_s: \pi_{s}=\pi_{s|F} \times \pi_F, \pi_s \text{ is supported on } A^{TI}\cap A^{EM-NTAT},\quad \pi_{DF;s|F} =\pi_{DF|F}^*\Bigg\},
\end{split}
\end{equation}
where $\pi_{DF|F}^*$ is the econometrician's choice of the belief of proportion of defiers. For example, if the econometrician believes that defiers are very unlikely to exists, then she can choose an exponentially decaying density function for $\pi_{DF|F}^*$. We use a single conditional marginal belief $\pi_{DF|F}^*$ in \eqref{eq: prior set}, but the econometrician may experiment with multiple beliefs to gain additional robustness, which will be discussed in Section \ref{section: General Theory}.


\subsubsection{The posterior distribution and interval estimates}
The econometrician observes the realized data $\bm{X}_n=\{(Y_i,D_i,Z_i)\}_{i=1}^n$ and uses the Bayes' rule to update her belief about the real data distribution to be $\pi_{F|\bm{X}_n}$. Because of the decomposition \eqref{eq: decomposition of prior believes}, the set of posterior is given by \begin{equation}\label{eq: posterior set}
\begin{split}
\Pi_{s|\bm{X}_n}=\Bigg\{\pi_s: \pi_{s}=\pi_{s|F} \times \pi_{F|\bm{X}_n},\quad \text{and}\quad  \pi_{DF;s|F}=\pi_{DF|F}^*\Bigg\}.
\end{split}
\end{equation}
The restrictions in the posterior set (\ref{eq: posterior set}) are similar to those in the prior set (\ref{eq: prior set}) except the belief of the data distribution is updated to $\pi_{F|\bm{X}_n}$. 

We propose the posterior upper- and lower-bound of the $LATE$ quantity:
\begin{equation}\label{eq: bound for LATE}
\begin{split}
LATE^*=\sup_{\pi_{s}\in \Pi_{s|\bm{X}_n}} \int_{\mathcal{S}} E_{G^s}[Y_i(1)-Y_i(0)|D_i(1)=1,D_i(0)=0] d\pi_{s},\\
LATE_*=\inf_{\pi_{s}\in \Pi_{s|\bm{X}_n}} \int_{\mathcal{S}} E_{G^s}[Y_i(1)-Y_i(0)|D_i(1)=1,D_i(0)=0] d\pi_{s},\\
\end{split}
\end{equation}
and the following target of the confidence set:
\begin{equation}\label{eq: confidence set for LATE}
\begin{split}
\sup_{\pi_{s}\in \Pi_{s|\bm{X}_n}}  Pr_{\pi_{s|\bm{X}_n}}	(LATE\in CI)\ge 1-\alpha,
\end{split}
\end{equation}
which is a uniform confidence region for all possible posterior beliefs. For any value $\widehat{LATE}\in [LATE_*,LATE^*]$, we can find a posterior ${\pi_{s}\in \Pi_{s|\bm{X}_n}}$ such that $\widehat{LATE}$ is the posterior mean corresponding to $\pi_s$.

\subsubsection{Characterization of the posterior quantities}
For a given $\pi_s$, there is no closed-form expression for the $LATE^*$, $LATE_*$ and $CI$. The computation method in \cite{giacomini2021robust} is developed for a finite-dimensional model and it can be hard to implement because the posterior $\pi_{s|\bm{X}_n}$ is a distribution over infinite dimensional objects. In this application, we use a combintation of analytical results and simulation to derive a feasible calculation of the bound $[LATE_*,LATE^*]$.

For any $\pi_{s|\bm{X}_n}\in \Pi_{s|\bm{X}_n}$, it can be written as the $\pi_{s|\bm{X}_n}=\pi_{s|F} \times \pi_{F|\bm{X}_n}$ for the uniquely chosen $\pi_{F|\bm{X}_n}$. Given the posterior for the data distribution $\pi_{F|\bm{X}_n}$, the variation in the posterior set  $\Pi_{s|\bm{X}_n}$ comes from the variation in $\pi_{s|F}$, which is the `non-updatable' part. There are existing nonparametric Bayesian methods for us to find $\pi_{F|\bm{X}_n}$ and we can simulate $F$ from $\pi_{F|\bm{X}_n}$. We then use the potential outcome framework to analytically characterize the conditional distribution $\pi_{s|F}$ that achieves the upper and lower bound. The following Lemma states the feasibility of separating the simulation and analytic analysis parts.

\begin{prop}\label{lem: separation of s|F and F}
Suppose $-\infty<LATE_*\le LATE^*<\infty$, then the following equalities hold:
	\begin{equation}\label{eq: $LATE$ bound in integraton form}
	\begin{split}
		LATE^*=\int_{\mathcal{F}}\int_{\mathbb{R}^+} \overline{LATE}(F,m) d\pi^*_{DF|F}(m) d\pi_{F|\bm{X}_n},\\
		LATE_*=\int_{\mathcal{F}}\int_{\mathbb{R}^+} \underline{LATE}(F,m) d\pi^*_{DF|F}(m) d\pi_{F|\bm{X}_n},
		\end{split}
	\end{equation}
	where
	\begin{equation}\label{eq: $LATE$ bound fixing F and m}
		\begin{split}
			\overline{LATE}(F,m)&=\sup_{s\in A': m^{df}(G^s)=m, M(G^s)=F} E_{G^s}[Y_i(1)-Y_i(0)|D_i(1)=1,D_i(0)=0],\\
			\underline{LATE}(F,m)&=\inf_{s\in A': m^{df}(G^s)=m, M(G^s)=F} E_{G^s}[Y_i(1)-Y_i(0)|D_i(1)=1,D_i(0)=0],
		\end{split}
	\end{equation}
	where $A'=A^{TI}\cap A^{EM-ATNT}$. 
\end{prop}	
Proposition \ref{lem: separation of s|F and F} separates the calculation of the $LATE$ bound into simulation part \eqref{eq: $LATE$ bound in integraton form} and the bound \eqref{eq: $LATE$ bound fixing F and m}. The bound \eqref{eq: $LATE$ bound fixing F and m} requires us to find the structure $s$ that maximizes/minimizes the $LATE$ quantity given a fixed number of defiers and a fixed observed data disribution.  In other words, for each possible value of the defier amount $m$, we find the most/least favorable structure $s$ and characterize the $LATE$ under this $s$. 

We briefly discuss the intuition behind Proposition \ref{lem: separation of s|F and F}. First, since we use a single $\pi_F$ and a single $\pi_{DF|F}^*$ to construct $\Pi_s$, by furthur conditioning on the value $m$, we have the decomposition $\pi_{s|\bm{X}_n}=\pi_{s|m,F}\times \pi^*_{DF|F}(m)\times \pi_{F|\bm{X}_n}$, which allows us to integrate over $\pi_{F|\bm{X}_n}$ and $\pi^*_{DF|F}$  in separate steps. It remains to optimize over $\pi_{s|m,F}$, which is equivalent to the pointwise optimization for each value of $m$ and $F$ as in \eqref{eq: $LATE$ bound fixing F and m} .

%

It remains to characterize the optimization over $\{s: m^{df}(G^s)=m, M(G^s)=F\}$, which is hard to find since it is still an infinite dimensional optimization problem. We take a final step to transform the problem into a tractable finite-dimensional optimization problem. 

\begin{thm}\label{thm: Transform infinite optimization to finite optimization}
	Let $p_F(y,d)$ and $q_F(y,d)$ be the densities of $P_F(\cdot,d)$ and $Q_F(\cdot,d)$ with respect to the dominating measure $\mu_F$.  We can compute $\overline{LATE}(F,m)$ using a 2-variable optimization program:
	\begin{equation}\label{eq: $LATE$ upper bound with fixed m}
	\overline{LATE}(F,m)=\sup_{a,b}\left[ \frac{\int_{\mathcal{Y}} y h_1^{max}(y)dy}{a} - \frac{\int_{\mathcal{Y}} yh_0^{max}(y)dy}{b} \right]
	\end{equation}
	subject to 
	\begin{equation}\label{eq: linear constraints on the conditional defiers}
	\begin{split}
	&\quad aPr_F(Z_i=0)+Pr_F(D_i=1)-Pr_F(D_i=1|Z_i=1)Pr_F(Z_i=0)\\
	&+bPr_F(Z_i=1)+Pr_F(D_i=0)-Pr_F(D_i=0|Z_i=0)Pr_F(Z_i=1)=m;\\
	&\int_{\mathcal{Y}} \max\{p_F(y,1)-q_F(y,1),0\}dy \le a\le \int_{\mathcal{Y}} p_F(y,1) dy;\\
	&\int_{\mathcal{Y}} \max\{q_F(y,0)-p_F(y,0),0\}dy \le b \le \int_{\mathcal{Y}} q_F(y,0)dy.
	\end{split}
	\end{equation}
	where $h_1^{max}$ and $h_0^{max}$ are specified in the following:
	\begin{equation}\label{eq: choice of h_1^max and h_0^max}
	\begin{split}
	h_1^{max}(y)=\max\{p_F(y,1)-q_F(y,1),0\} + \min\{p_F(y,1),q_F(y,1)\}\mathbbm{1}(y> \bar{y}_{max})\\
	h_0^{max}(y)= \max\{q_F(y,0)-p_F(y,0),0\}+ \min\{q_F(y,0),p_F(y,0)\}\mathbbm{1}(y< \underline{y}_{max})
	\end{split}
	\end{equation}
	with $\bar{y}_{max}$ and $\underline{y}_{max}$ be the solution to 
	\[
	\begin{split}
	\int_{\bar{y}_{max}}^{\infty} \min\{p_F(y,1),q_F(y,1)\} dy= a- \int_{\mathcal{Y}} \max\{p_F(y,1)-q_F(y,1),0\}dy\\
	\int^{\underline{y}_{max}}_{-\infty}  \min\{q_F(y,0),p_F(y,0)\} dy= b- \int_{\mathcal{Y}} \max\{q_F(y,0)-p_F(y,0),0\}dy.
	\end{split}
	\]
	
	Similarly,  
	\[
	\underline{LATE}(F,m)=\inf_{a,b}\left[ \frac{\int_{\mathcal{Y}} y h_1^{min}(y)dy}{a} - \frac{\int_{\mathcal{Y}} yh_0^{min}(y)dy}{b} \right]
	\]
	subject to the same linear constraint (\ref{eq: linear constraints on the conditional defiers}).	The $h_1^{max}$ and $h_0^{max}$ are specified in the following:
	\[
	\begin{split}
	h_1^{min}(y)=\max\{p_F(y,1)-q_F(y,1),0\} + \min\{p_F(y,1),q_F(y,1)\}\mathbbm{1}(y< \bar{y}_{min})\\
	h_0^{min}(y)= \max\{q_F(y,0)-p_F(y,0),0\}+ \min\{q_F(y,0),p_F(y,0)\}\mathbbm{1}(y> \underline{y}_{min})
	\end{split}
	\]
	with $\bar{y}_{min}$ and $\underline{y}_{min}$ be the solution to
	\[
	\begin{split}
	\int^{\bar{y}_{min}}_{-\infty}  \min\{p_F(y,1),q_F(y,1)\} dy= a- \int_{\mathcal{Y}} \max\{p_F(y,1)-q_F(y,1),0\}dy\\
	\int_{\underline{y}_{min}}^{\infty} \min\{q_F(y,0),p_F(y,0)\} dy= b- \int_{\mathcal{Y}} \max\{q_F(y,0)-p_F(y,0),0\}dy.
	\end{split}
	\]
\end{thm}
The optimization problem \eqref{eq: $LATE$ upper bound with fixed m}-\eqref{eq: linear constraints on the conditional defiers} optimize over $a$ and $b$ and the constraints \eqref{eq: linear constraints on the conditional defiers}  are  in linear form can be implimented using standard optimization packages. 

We provide some intuition for Theorem \ref{thm: Transform infinite optimization to finite optimization}. Let $h_1(y)$ denotes the density of $Y_i(1)$ and compliers conditional on $Z_i=1$,\footnote{That is, $h_1(y)=d Pr(Y_i(1)\le y,D_i(1)=1,D_i(0)=0|Z_i=1)/d\mu_F$.}  and let $h_0(y)$ denotes the density of $Y_i(0)$ and compliers conditional $Z_i=0$. By definition of $LATE$, for this $h_1(y)$ and $h_0(y)$, we have $$LATE=\frac{\int_{\mathcal{Y}} yh_1(y)dy}{ \int_{\mathcal{Y}} h_1(y)dy} -\frac{ \int_{\mathcal{Y}} yh_0(y)dy}{ \int_{\mathcal{Y}} h_0(y)dy}.$$ At the same time, the amount of compliers, conditional on $Z_i=1$, is $\int_{\mathcal{Y}} h_1(y)dy$. So essentially,  the optimization \eqref{eq: $LATE$ upper bound with fixed m} is converted to a two step optimization: 1.Given the value of compliers amounts for different $Z_i$ values (i.e., $a$ and $b$), what is the optimal densities; 2. What is the optimal $a$ and $b$ value.

The first question is answered by the $h_1^{max}(y)$ and $h_0^{min}(y)$. Fixing the $a$, to maximize $\int_{\mathcal{Y}} yh_1(y)dy/a$, we must allocate more probability masses to larger values of $y$. However, we also face the testable implication \eqref{eq: testable implication of LATE} in the density form, and there is a minimum amount of density ($\max\{p_F(y,1)-q_F(y,1),0\}$) we must allocate to a $y$ value. In view of (\ref{eq: choice of h_1^max and h_0^max}), the first additive term of $h_1^{max}$ reflects this minimal level. After maintaining this essential density of compliers, we are left with $a- \int_{\mathcal{Y}} \max\{p_F(y,1)-q_F(y,1),0\}dy$ amount of compliers, which should be allocated to all values above the $\bar{y}_{max}$. At the same time, we cannot allocate too much density to large value of $h_1^{max}$, otherwise the testable implication for always takers will kick in, which is reflected in the second additive term of $h_1^{max}$ in (\ref{eq: choice of h_1^max and h_0^max}).\footnote{The maximal density for compliers conditional $Z_i=1$ is $p_F(y,1)$, and the difference  between $\max\{p_F(y,1)-q_F(y,1),0\}$ and $p_F(y,1)$ is $\min\{p_F(y,1),q_F(y,1)\}$, which is the room for us to manipulate.} The rationale for $h_0^{max}$ is the same and we allocate all masses to small values of $y$. Given the value of $a$ and $b$, we can calculate $h_{1}^{max}$ and $h_{0}^{max}$ numerically. 

The optimization of $a$ and $b$ does not have a closed-form solution and we have to use numerical optimization package.  The first constraint in \eqref{eq: linear constraints on the conditional defiers} reflects that we have a fixed amount of defiers $m$, and the constraint is constructed via the potential outcome equation \eqref{eq: potential outcome}. The second and third constraints come from the testable implication \eqref{eq: testable implication of LATE}.

Combining Proposition \ref{lem: separation of s|F and F} and Theorem \ref{thm: Transform infinite optimization to finite optimization}, we have a way to calculate the $LATE^*$, $LATE_*$ and the confidence set. We first generate a sample of $F$ from $\pi_{F|\bm{X}_n}$, and for each realized $F$, we generate a $m$ value from $\pi_{DF|F}^*$. Given the pair $(F,m)$, use Theorem \ref{thm: Transform infinite optimization to finite optimization} to generate a sample of $\overline{LATE}(F,m)$ and $\underline{LATE}(F,m)$. The $LATE^*$ and $LATE_*$ are the simulated sample mean of $\overline{LATE}(F,m)$ and $\underline{LATE}(F,m)$ correspondingly.  The confidence interval can also be construct from the distribution of $\overline{LATE}(F,m)$ and $\underline{LATE}(F,m)$ under  $\pi_{F|\bm{X}_n}\times \pi_{DF|F}^*$. Let $\underline{C}_{\alpha/2}$ be the $\alpha/2$ quantile of $\underline{LATE}(F,m)$ under $\pi_{F|\bm{X}_n}\times \pi_{DF|F}^*$, and let  $\bar{C}_{1-\alpha/2}$ be the $1-\alpha/2$ quantile of $\overline{LATE}(F,m)$ under $\pi_{F|\bm{X}_n}\times \pi_{DF|F}^*$. We use $[\underline{C}_{\alpha/2}, \bar{C}_{1-\alpha/2}]$ as the confidence interval.

\begin{prop}\label{prop: confidence set consistent, LATE}
	The interval $CI\equiv[\underline{C}_{\alpha/2}, \bar{C}_{1-\alpha/2}]$ is a valid confidence interval in regard to (\ref{eq: confidence set for LATE}).
\end{prop}
In Proposition \ref{prop: confidence set consistent, LATE}, we do not use the $\alpha$-quantile because we want to get some robustness when the upper and lower bound are very close to each other. However, when the posterior mean bounds are wide, the confidence interval can be conservative.\footnote{Additional moment-selection based \citep{andrews2010inference} method can be adapted in the Bayesian inference context, but we leave it to keep the topic concentrated.}

\subsection{Frequentist Equivalence} \label{section: frequentist Equivalence }
We now derive an equivalence result of our method to a frequentist approach to the refutable models. More equivalence results can be found in Section \ref{section: General Theory}. In the Bayesian statistics literature, the Berstein-von Mises theorem characterizes the equivalence between the Bayesian posterior distribution and the limit distribution under the frequentist world. In other words, Bayesian \textit{inference} is equivalent to frequentist \textit{inference}.

We do not seek to characterize the equivalence in the spirit of the Berstein-Von Mises theorem. This is because the structural model here is an infinite-dimensional and the nonparametric version of the Berstein-von Mises theorem is only valid under very specific models. Instead, we aim to show that the $LATE$ estimator under different frequentist models can be justified by a corresponding prior belief. As a result, we show that the frequentist \textit{assumption set} is equivalent to a Bayesian prior \textit{belief specification}. 

We start with the crucial assumption on nonparametric posterior convergence result, and briefly describe the frequentists' minimal relaxation method. 

\begin{assumption}\label{assumption: convergence of the posterior for $LATE$ application}
	Let $d_\mathcal{F}$ be the total variation metric on the $\mathcal{F}$. The posterior $\pi_{F|\bm{X}_n}$ converges weakly to the degenerate distribution with a point mass at $F_0$ along the empirical distribution sequence $\mathbb{P}_{F_0}^{(n)}$.
\end{assumption}

\begin{definition}\label{def: minimal deviation belief}
	Let $m_{min}^{df}(F)=\min_{s\in A^{TI}\cap A^{EM-NTAT}: F= M(G^s)} m^{df}(s)$ be the minimal probability of defiers that is required to rationalize $F$.
	We say a prior belief set $\Pi_s$ on $A^{TI}\cap A^{EM-NTAT}$ satisfies the minimal defier constraints if $\pi_{DF|F}^*= \delta(m_{min}^{df}(F))$, where $\delta(x)$ denotes the point mass measure at $x$. 
\end{definition}

\begin{prop}\label{prop: Frequentist to minimal deviation}
	Suppose Assumption \ref{assumption: convergence of the posterior for $LATE$ application} holds and $\mathcal{Y}$ is bounded. Let $\Pi_s^{Min}$ denote the prior set that satisfy the minimal defier constraints. 
	Let $(Y_i,D_i,Z_i)\sim F_0$ be the true distribution of data, then the robust Bayesian estimator satisfies:
	\begin{equation}\label{eq: convergence to the frequentist LATE}
	LATE^*=LATE_*\rightarrow_p \frac{\int_{\mathcal{Y}_1(F_0)} y(p_{F_0}(y,1)-q_{F_0}(y,1)) dy}{\int_{\mathcal{Y}_1(F_0)} (p_{F_0}(y,1)-q_{F_0}(y,1)dy} - \frac{\int_{\mathcal{Y}_0(F_0)} y(q_{F_0}(y,0)-p_{F_0}(y,0)) dy}{\int_{\mathcal{Y}_0(F_0)} (q_{F_0}(y,0)-p_{F_0}(y,0)dy},
	\end{equation}
	where $\mathcal{Y}_1(F_0)=\{y\in\mathcal{Y}: p_{F_0}(y,1)\ge q_{F_0}(y,1)\}$ and $\mathcal{Y}_0(F_0)=\{y\in\mathcal{Y}: q_{F_0}(y,0)\ge p_{F_0}(y,0)\}$.
\end{prop}
The limit in \eqref{eq: convergence to the frequentist LATE} is the frequentists' minimal deviation method identified $LATE$ \citep{liao2020estimating,dahl2023nevertoolate}. Proposition \ref{prop: Frequentist to minimal deviation} shows that the interval $ [LATE_*, LATE^*]$ shrinks to a point and converges to the minimal-deviation characterized identified quantity \eqref{eq: convergence to the frequentist LATE}. The result shows that the minimal-deviation method is equivalent to imposing a degenerate belief on the deviation from the `No Defiers' assumption. We use the Dirichlet process to model the prior distribution because it is known to have a consistenty result and computationally feasible posterior. We delegate the background of the Dirichlet process to \cite{ghosal2017fundamentals} for readers who are not familiar with the nonparametric Bayes method, especially Chapters 2-4.

\subsection{Posterior via Dirichlet Mixture Process Prior}
We now propose a Bayesian estimation and inference procedure based on simulation. First, we specify a Dirichlet mixture process for the prior belief. The Dirichlet mixture process can accommodate density-based estimation than the Dirichlet process alone.

\paragraph{The prior of $F$} Weconsider a hierarchical prior to separate the discrete and the continuous part of $F$:
\begin{eqnarray}
\pi_{dz}\sim \text{Dirichlet} (4,H_{dz}),\label{eq: prior for dz}\\
\pi_{\theta|dz}\sim  \text{Dir} (H_{\theta|dz})\label{eq: prior for intermediate parameter},\,\, \sigma_{dz}^{-2}\sim \pi_\sigma\equiv  Gamma(\alpha,\beta), \label{eq: conditional prior for theta and sigma} \\
\pi_{y|dz}\sim \int_{\sigma}\int_{\theta} \psi_{\sigma_{dz}}(y;\theta)d \pi_{\theta|dz} d \pi_\sigma\label{eq: prior for y|dz},\\
(Y_i,D_i,Z_i) \sim \pi_{dz}\times \pi_{y|dz}\label{eq: distribution for ydz}.
\end{eqnarray}
We briefly discuss the notation in (\ref{eq: prior for dz})-(\ref{eq: distribution for ydz}) by comparing the notations to the parametric Bayesian method. We specify the marginal prior belief of $(D_i,Z_i)$ to satisfy a Dirichlet distribution with scale parameter $\alpha_{dz}$ and centering distribution $H_{dz}$. Here, $H_{dz}$ is a discrete measure supported on $\{0,1\}\times \{0,1\}$, which serves as the center of the Dirichlet distribution. The number $4$ specifies that $H_{dz}$ is supported on 4 discrete points.  The finite measure $H_{dz}$ does not have to be a probability measure, but its total mass $|H_{dz}|\equiv H_{dz}(D_i\in\{0,1\},Z_i\in\{0,1\})$ is related to `variance' of the random draw. Roughly speaking, if $|H_{dz}|$ is larger, the prior belief weights more in the posterior. Any distribution drawn from $\text{Dirichlet} (4,H_{dz})$ can be viewed as a perturbation of the distribution $H_{dz}/|H_{dz}|$.

We use a different notation $\text{Dir}(\cdot)$ to denote the Dirichlet process. For any finite number of partitions of $\mathcal{Y}$, denoted by $\mathcal{Y}_1,...,\mathcal{Y}_K$, the finite discrete distribution generated from the Dirichlet process follows a Dirichlet distribution: $(\pi_{\theta|dz}(\mathcal{Y}_k)))_{k=1}^K\sim \text{Dirichlet}(K,(H_{\theta|dz}(\mathcal{Y}_k))_{k=1}^K)$. In other words, the Dirichlet process is the Kolmogrove-extension of the Dirichlet distribution.  We specify the conditional prior belief of $Y_i|D_i,Z_i$ distribution as a Dirichlet process mixture as in (\ref{eq: prior for y|dz}), which can be viewed as the weighted sum of densities for at different $\theta$ locations. Here $\psi_\sigma(y,\theta)$ is a kernel density function with auxiliary parameter $\sigma$. The parameter $\theta$ is assumed to follow a further Dirichlet process $\text{Dir} (H_{\theta|dz})$. The $Gamma(\alpha,\beta)$ distribution is chosen to be independent of the $Dir(H_{\theta|dz})$ process.\footnote{The additional parameter $\sigma$ serves as the mixture of Dirichlet Process Mixture. The additional mixture facilitates computation in many applications.} Depending on the application, $\psi$ can be chosen to be a normal pdf with mean $\theta$ and standard deviation $\sigma$ if $y$ is univariate and unbounded; When $y$ is bounded, $\psi_\sigma(y;\theta)$ can be chosen to be $Beta(a,b)$ function, where we choose $\alpha=0$\footnote{The choice of $\alpha=0$ is specific to the Beta kernel function and cannot be used for the Gaussian kernel.} so that $\sigma\equiv 0$ is a degenerate parameter, and $a<1\le b$ is sampled from $\theta=(a,b)\sim\pi_{\theta|dz}$.  

\paragraph{Data-updated posterior}  Recall that $\bm{X}_n=\{(Y_i,D_i,Z_i)\}_{i=1}^n$ denotes the observed data and let $\mathbb{F}^{DZ}_n$ denote the empirical distribution of data. The posterior $\pi_{dz|\bm{X}_n}$ of the distribution of $(D_i,Z_i)$ follows a Dirichlet distribution with parameter $(4, \frac{|H_{dz}|}{|H_{dz}|+n}H_{dz}+ \frac{n}{|H_{dz}|+n}\mathbb{F}^{DZ}_n)$. To calculate the conditional posterior $\pi_{y|D_i=d,Z_i=z,\bm{X}_n}$, we need to get the posterior of $\text{Dir} (H_{\theta|dz}|\bm{X}_n)$ and the posterior of $\Gamma(\alpha,\beta|\bm{X}_n)$. For the calculation of $\pi_{y|D_i=d,Z_i=z,\bm{X}_n}$, we will only use the observation with $D_i=d$ and $Z_i=z$. However, the conditional posterior $\pi_{y|D_i=d,Z_i=z,\bm{X}_n}$ does not have a closed-form solution but it can be calculated using the MCMC method, and an existing \texttt{R} package \texttt{dirichletprocess} can be used directly. 

We will focus on the normal kernel $\psi$. The joint posterior $\pi_{F|\bm{X}_n}$ is known to satisfy the consistency Assumption \ref{assumption: convergence of the posterior for $LATE$ application} under mild conditions.

\begin{assumption}\label{assumption: consistency requirement for normal kernel}
	The true densities $p_{F_0}(y,d)$ and $q_{F_0}(y,d)$ satisfy the following entropy constraints for $d\in\{0,1\}$:\begin{enumerate}
		\item $\int_{\mathcal{Y}} p_{F_0}(y,d) \log p_{F_0}(y,d) dy<\infty$ and $\int_{\mathcal{Y}}  q_{F_0}(y,d) \log q_{F_0}(y,d) dy<\infty$;
		\item $-\int_{\mathcal{Y}}  p_{F_0}(y,d) \log\left[\inf_{||y'||<\delta}p_{F_0}(y-y',d) \right] dy<\infty $, and \\$-\int_{\mathcal{Y}}  q_{F_0}(y,d) \log\left[\inf_{||y'||<\delta}q_{F_0}(y-y',d) \right] dy<\infty $ for some $\delta>0$. 
	\end{enumerate}
Moreover, the parameter $\alpha>1$ in the Gamma distribution.
\end{assumption}

\begin{prop}\label{prop: consistent posterior}
	Assumption \ref{assumption: consistency requirement for normal kernel} implies that the posterior estiamted under (\ref{eq: prior for dz})-(\ref{eq: distribution for ydz}) satisfies Assumption \ref{assumption: convergence of the posterior for $LATE$ application} for the normal kernel $\psi_{\sigma}(\cdot;\theta)$. 
\end{prop}
Proposition \ref{prop: consistent posterior} establish the posterior consistency result and also validate the frequentist equivalence result in Proposition \ref{prop: Frequentist to minimal deviation}. We conclude the section by summarizing the poterior sampling altorithm.

\begin{algorithm}[H]
	\caption{Posterior Sampling Algorithm.}
	Combining Proposition \ref{lem: separation of s|F and F} and Theorem \ref{thm: Transform infinite optimization to finite optimization}, we have a computationally feasible algorithm to calculate the ${LATE}^*$, ${LATE}_*$ in (\ref{eq: bound for LATE}) and the confidence set $CI$ in (\ref{eq: confidence set for LATE}):
	\begin{enumerate}
		\item Setup the Dirichlet prior for the distribution of observed variable $\pi_F$ as in  (\ref{eq: prior for dz})-(\ref{eq: distribution for ydz}),  and set up a marginal distribution for defiers $\pi_{DF|F}^*$ for each F.
		\item Use nonparametric Bayesian method to update the distribution of $F$ conditioning on the data observation $\bm{X}_n$. Denote the posterior as $\pi_{F|\bm{X}_n}$.
		\item Draw a realization $F$ from the posterior $\pi_{F|\bm{X}_n}$ via MCMC method. The \texttt{R} package \texttt{dirichletprocess} is used here. 
		\item Given the $F$, randomly sample an $m$ from the conditional distribution  $\pi_{DF|F}^*$. Calculate $\overline{LATE}(F,m)$ and $\underline{LATE}(F,m)$ using the optimization problem in Theorem \ref{thm: Transform infinite optimization to finite optimization}.
		\item Repeat steps 3 and 4 for different draws of $F$ which generates a sample of  $\overline{LATE}(F,m)$ and $\underline{LATE}(F,m)$. Let $LATE^*$ be the sample mean of $\overline{LATE}(F)$, and let $\underline{LATE}(F,m)$ be the sample mean of $\underline{LATE}(F,m)$. Let $CI=[\underline{C}_{\alpha/2}, \bar{C}_{1-\alpha/2}]$, where $\underline{C}_{\alpha/2}$ is the $\alpha/2$ quantile of  $\underline{LATE}(F,m)$ and $\bar{C}_{1-\alpha/2}$ is the $1-\alpha/2$ quantile of $\overline{LATE}(F,m)$. 
	\end{enumerate}
\end{algorithm}

\subsection{Empirical Application }

In this section, we apply the robust Bayesian method to \citet{card1993using}, who studied the causal effect of college attendance on earnings. We take the outcome variable $Y_i$ to be an individual $i$'s log wage in 1976, and  $D_i=1$ to be individual $i$'s four-year college attendance. We use the college proximity variable $Z_i$ as an instrument for college attendance, i.e. $Z_i=1$ means the individual was born near a four-year college. The empirical setting has been used by both \citet{kitagawa2015} and \citet{mourifie2017testing} to test the IA-M assumption. In various settings, the IA-M assumption is rejected, and it is reasonable to believe that defiers are likely to present. 

We follow  \citet{mourifie2017testing} to condition $(Y_i,D_i,Z_i)$ on three characteristics: living in the south (S/NS), living in a metropolitan area (M/NM), and an African-American ethnic group (B/NB). We also drop the NS/NM/B and NS/M/B group due to a small sample size or a small number of $Z_i=0$. 

For the choice of prior parameters in (\ref{eq: prior for dz})-(\ref{eq: distribution for ydz}), we choose $H_{dz}$ to be the uniform distribution over $\{0,1\}\times \{0,1\}$, and $\psi(\cdot;\theta)$ to be normal density function with $\theta=(\mu,\sigma)$. The centering measure $H_{(\mu,\sigma)|dz}$ is assumed to follow the standard bivariate normal distribution. We choose the parameter $(\alpha,\beta)=(2,4)$ which is the default value in the \texttt{R} package. We choose $\pi_{DF|F}^*$ to be a Gaussian-decaying density $\pi_{DF|F}^*(m)=\frac{C}{\sqrt{2\pi}\sigma(F)}e^{-(m-m^{min}(F))^2/\sigma^2(F)}$ on  $[m^{min}(F),m^{max}(F)]$, zero otherwise, where $m^{min}(F)$ and $m^{max}(F)$ are correspondingly the minimal and maximal amount of defiers required to justify the $F$ distribution\footnote{That is $m^{min}(F)=\min\{m^{df}(s): F=M(G^s)\}$. Similar definition holds for $m^{max}(F)$.}. We choose $\sigma(F)=(m^{max}(F)-m^{min}(F))/1.96$. The constant $C$ is chosen to make $\pi_{DF|F}^*$ a proper density function. The estimation and inference results are shown in Table \ref{table: estimation and inference for LATE}.
{
	\begin{table}[h]
		\renewcommand{\arraystretch}{1.5}
		\footnotesize
		
		\caption{Estimation Result with Robust Bayesian Method }
		\scalebox{0.9}{
			
				\begin{tabular}{lcccccc}\label{table: estimation and inference for LATE}
				Group                     & NS,NM,NB           & NS,M,NB            & S,NM,NB            & S,NM,B            & S,M,NB             & S,M,B                            \\ 
				Observations              & 429                & 1191               & 307                & 314               & 380                & 246                         \\ \hline
				$[LATE_*,LATE^*]$         & {[}-0.61,  3.58{]} & {[}0.70,  3.07{]}  & {[}-0.69,  2.64{]} & {[}0.25,  2.94{]} & {[}0.60,  3.37{]}  & {[}0.64,  2.67{]}   \\
				CI                      & {[}-4.77,  7.94{]} & {[}-2.18,  4.40{]} & {[}-4.32,  5.97{]} & {[}-3.01, 5.25{]} & {[}-2.16,  4.99{]} & {[}-2.63, 4.42{]} \\ \hline
				Frequentist Minimal                                                    & 0.5599           & 0.1546               & 0.2524             & 0.4773             & 0.5276               & 0.4358                     \\
				CI for $LATE^{ID}_{\tilde{A}}(F)$                                                & {[}0.01, 1.11{]} & {[}-0.51, 0.82{]} & {[}-1.22,	1.73{]} & {[}-0.09,1.04{]} & {[}-2.54,	3.59{]}     & {[}-5.15,	6.02{]}  \\\hline
				$LATE^{wald}(F)$          & 0.5976             & 0.0761             & -6.4251            & 1.1873            & -1.5412            & 17.9620                       \\
				CI for ${LATE}^{wald}(F)$ & {[}-0.20,1.39{]}    & {[}-1.24,1.39{]}   & {[}-105,92{]}      & {[}-0.53,2.90{]}  & {[}-5.09,2.01{]}   & {[}-1.7e4,1.7e4{]}  \\\hline
			\end{tabular}
		}
\end{table}}

There are several interesting observations from Table \ref{table: estimation and inference for LATE}. First, the robust Bayesian interval estimate is more stable than the $LATE^{wald}$. The $LATE^{wald}$ can estimate very negative $LATE$ (S/NM/NB and S/M/NB groups) or unrealistic high $LATE$ (S/M/B). In comparison, the set estimator under the robust Bayesian method is more persistent across different groups.  We also compare the result under the robust Bayesian method to that under the frequentist minimal defiers, see Proposition \ref{prop: Frequentist to minimal deviation}. The set estimator and the confidence set under the robust Bayesian method are wide. Therefore, frequentists' identified set under the minimal-defier Assumption (see Proposition \ref{prop: Frequentist to minimal deviation}) may generate over-precise $LATE$ estimates. The wider $LATE$ bound in the robust Bayesian framework shows that the quantity is very sensitive to the presence of defiers, and the frequentist minimal defiers approach can generate over-confident interpretation of the return of college education.

\section{A General Theory}\label{section: General Theory}
In this section, we develop a robust Bayesian theory for models with refutable assumptions that generalizes the insight from the $LATE$ application to incomplete models. We continue using the notations in the $LATE$ model and revisit the $LATE$ application whenever a new concept is introduced. We delegate further illustrating examples to Section \ref{sec: Additional Applications}.

\subsection{Definitions}
Starting from a vector of observed variables $X$, we can define the observation space $\mathcal{F}$ as the collection of all regular distributions of $F(X)$. The regularity condition, such as the smoothness of $F(X)$ or existence of certain moments of $X$, are conviently chosen by the econometrician to analyze the model.\footnote{These conditions, such as exitence of moments or densities, are often non-testable given a finite sample.}

Following \citet{koopmans1950identification} and \citet{jovanovic1989}, we consider that any outcome $X$ is generated through some distribution of underlying random vector  $\epsilon$ through some mapping $M$.  A pair of a distribution of $\epsilon$, denoted by $G$, and a mapping $M$ is called an econometric structure. The distribution $G$ governs the fundamental heterogeneity at the observation level. The mapping $M$ governs the economic machenism that generates the observed data $X$ from $\epsilon$.  Since in most econometric problems, we focus on the distribution of outcomes $F$ instead of how each $X$ is related to $\epsilon$, we directly define the mapping $M$ as a mapping from the space of underlying variable distributions to $\mathcal{F}$.

\begin{definition}\label{def: econometric structure}
	An econometric structure $s=(G^s,M^s)$ consists of a distribution $G^s$ of $\epsilon$, and an outcome mapping $M^s$. Let $\mathcal{G}$ denote the space of all regular distributions of $G^s(\epsilon)$. The outcome mapping $M^s$ is a function $M^s:\mathcal{G}\rightarrow \mathcal{F}$. 
\end{definition}
The requirement of the regularity of $G^s$ is in the same spirit as that of $F$. Unlike in the $LATE$ example where $M$ is fixed for all strutures, the outcome mapping $M^s$ can also vary across econometric structures to capture differences in economic mechanisms. Think of an OLS regression $Y_i=\beta^s_0+\beta^s_1 X_i +\eta_i$. A structure consists of a distribution $G^s$ of $(X_i,\eta_i)$ and a mapping $M^s$ which can be characterized by $(\beta^s_0,\beta^s_1)$.

\begin{definition}\label{def: structure universe}
	A structure universe $\mathcal{S}$ is a collection of structures such that $\cup_{s\in\mathcal{S}} M^s(G^s)=\mathcal{F}$, and an assumption $A$ is a subset of $\mathcal{S}$. 
\end{definition}
The structure universe determines the paradigm that can encompass different contexts while $A$ imposes assumptions that are convenient for a particular empirical context. In view of the OLS regression where we believe the average effects of $X_i$ on $Y_i$ is positive, the $\mathcal{S}$ does not restrict the parameter space, but $A=\{s: \beta_1^s\ge 0\}$ is a restriction. An assumption $A$ is called refutable if there exists some observed distribution $F\in\mathcal{F}$ such that $F\notin\cup_{s\in A} M^s(G^s)$. In other words, we can find an observed data distribution that cannot be generated by econometric structures inside $A$. In this case, we are aware of the possible violation of $A$, and a prior belief that putting all probability on $A$ is inappropriate. Instead, we may consider that $A$ is violated, but the violation of $A$ is a rare event and correspondingly form a prior belief. 

To formalize the construction of the prior belief set, we first consider that the original assumption can be written as the intersection of countably many sub-assumptions $A=\cap_{j=1}^\infty A_j$, and for each assumption, we find a deviation metric $m_j(s):\mathcal{S}\rightarrow \mathbb{R}_+$ that measures the deviation of structure $s$ from $A_j$. Moreover, we require that the deviation metric satisfies
\[
A=\{s: m_j(s)=0\}\cap (\cap_{l\ne j} A_{l} ).
\]
This representation above requires that $m_j(s)$ is a sharp characterization of $A_j$ given the rest of the assumptions, which allows us to use $m_j(s)$ to describe our belief of deviation from the baseline model $A$. The choice of $A_j$, $m_j$ and the multiplicity of ways to relaxed assumption are discussed in \cite{liao2020estimating}.\footnote{{Also see Appendix \ref{sec: monotone IV + choice of m_j} for an additional example that illustrates the multiplicity issue.}}

\subsection{Robust Bayesian Prior}
The econometrician wants to take a stance on assumption $A_j$ while leaving holding all other assumptions untouched. To do so, she puts a prior $\pi_{s}$ over the rest of the assumptions $\cap_{l\ne j} A_{l}$. The prior belief $\pi_{s}$ induces a marginal distribution on $\mathcal{F}$ via 
\[
\pi_F (\mathcal{F}_0)=\int_{\mathcal{S}} \mathbbm{1}(M^s(G^s)\in \mathcal{F}_0) d\pi_{s}.
\]
Unlike finite dimensional models where metrics on econometric structures are natrual, when we extend the parametric analysis to nonparametric models, we need to further make topological assumptions on $\mathcal{S}$ and $\mathcal{F}$. We endow the structural space with a metric $d_s$ and the observed distribution space $\mathcal{F}$ with a metric $d_\mathcal{F}$. When $(\mathcal{S},d_s)$ and $(\mathcal{F},d_F)$ are polish spaces, conditional distribution is well-defined\footnote{See Chapter 2 of \cite{ghosal2017fundamentals}.}, and the prior belief $\pi_s$ can be characterized as the product of two measures
\[
\pi_s= \pi_{s|F}\times \pi_{F}.
\]
Since we want to maintain the rest of the assumptions, we assume that $\pi_s$ is supported on $\cap_{l\ne j} A_{l}$. The econometrician may not want to put a single $\pi_s$ because by doing so she also takes stances on the likelihood of other aspects of the model.\footnote{In the $LATE$ example, a single prior also specifies the distribution of always takers. In the OLS regression, putting a single prior not only specifies the likelihood of deviation from $\{s: \beta_1^s\ge0\}$ but also the distribution of $\beta_0^s$. } Instead, she believes that only $A_j$ is likely to be violated and is only willing to make a statement with respect to the distribution of $m_j(s)$. In this case, we propose a robust Bayesian prior set:
\begin{equation}\label{eq: general prior set}
\Pi_{s}=\{\pi_s: \pi_s=\pi_{s|F}\times \pi_F, \quad \pi_s \text{ supported on } \cap_{l\ne j} A_{l},\quad \pi_{m_j|F}\in \Pi_{m_j|F}\}.
\end{equation}
where $\pi_{m_j|F}$ is the marginal distribution of the deviation from $A_j$ conditional on the $F$, i.e. $\pi_{m_j|F}(B)\equiv \int_{\mathcal{S}} \mathbbm{1}(m_j(s)\in B) d \pi_{s|F} $ for any measurable set $B$. The general prior set (\ref{eq: general prior set}) is slightly different from \eqref{eq: prior set}: We allow the marginal belief of deviation from $A_j$ to fall in a general set $\Pi_{m_j|F}$ rather than choosing a single marginal belief.  The set $\Pi_{m_j|F}$ specifies the econometricians' belief of the deviation from the $A_j$ assumption and allows for additional robustness against the choice of prior.  The set of beliefs can change with $F$ to allow for flexibility of specification of beliefs. The following are some examples of the possible choices of $\Pi_{m_j|F}$:
\begin{equation}\label{eq: some example of prior sets}
	\begin{split}
		\Pi_{m_j|F}^1&=\{\pi_{m_j|F}: \text{density of } \pi_{m_j|F} \text{ is weakly decreasing on $[a^F_{min},a^F_{max}]$}\};\\
		\Pi_{m_j|F}^2&=\{ \text{density of } \pi_{m_j|F} \text{ is } t/(a^F_{max}-a^F_{min})\};\\
		\Pi_{m_j|F}^3&=\{ \text{density of } \pi_{m_j|F} \text{ is proportional to } e^{-t^2/\sigma^2} \text{ on $[a^F_{min},a^F_{max}]$} \};
	\end{split}
\end{equation}
where $a^F_{min}$ and $a^F_{max}$ are the $F$ induced minimal and maximal deviations.\footnote{In the LATE example, the constraints on $a$ and $b$ in \eqref{eq: linear constraints on the conditional defiers} set the bounds for probability of defiers.}
The first set is the collection of all possible prior beliefs that have a decreasing density of deviation, this is a large set and may impose too few restrictions for the model to generate informative results. The second one is a uniform prior while the third is a truncated normal density. We can also use convex combination of two sets to create a new prior set: $\Pi^{mix}_{m_j|F}=\alpha\Pi_{m_j|F}^2+(1-\alpha)\Pi_{m_j|F}^3$, for $\alpha\in[0,1]$.

\subsection{Characterization of the posterior of parameters}
The econometrician is interested in an 1-dimensional parameter $\theta$, which is a function from the structure space to the parameter space $\theta(s): \mathcal{S} \rightarrow \Theta \subseteq \mathbb{R}$. 

After observing data realization $\bm{X}_n$, the econometrician's posterior belief set is given by  
\[
\Pi_{s|\bm{X}_n}=\{\pi_{s|\bm{X}_n}:\pi_{s|\bm{X}_n}=\pi_{s|F}\times \pi_{F|\bm{X}_n}, \quad \pi_{s|\bm{X}_n}\text{ supported on } \cap_{l\ne j} A_{l},\quad \pi_{m_j|F}\in \Pi_{m_j|F}\} 
\]
The above posterior set shows the observed data only updates the belief of data distribution $F$, and it leaves the conditioning belief $\pi_{s|F}$ unchanged. The posterior beliefs set then induce a set of beliefs of the parameter of interest:
\begin{equation}\label{eq: general posterior set}
\Pi_{\theta|\bm{X}_n} =\left\{ \pi_{\theta|\bm{X}_n}: \pi_{\theta|\bm{X}_n}(B)= \int_{\mathcal{S}} \mathbbm{1}(\theta(s) \in B) d\pi_{s|\bm{X}_n}, \quad \text{ and }\pi_{s|\bm{X}_n}\in \Pi_{s|\bm{X}_n}\right\}.
\end{equation}
The corresponding bound estimator of the parameter of interest is given by 
\[
\theta^*=\sup_{\pi_{\theta|\bm{X}_n}\in \Pi_{\theta|\bm{X}_n}}E_{\pi_{\theta|\bm{X}_n}}[\theta],\quad  \theta_*=\inf_{\pi_{\theta|\bm{X}_n}\in \Pi_{\theta|\bm{X}_n}}E_{\pi_{\theta|\bm{X}_n}}[\theta].
\]
For inference, we seek a confidence set $CI$ such that 
\[
\inf_{\pi_{\theta|\bm{X}_n}\in \Pi_{\theta|\bm{X}_n}} Pr_{\pi_{\theta|\bm{X}_n}} (\theta\in CI)\ge 1-\alpha.
\]
The optimization with respect to the $\pi_{\theta|\bm{X}_n}$ is an infinite dimensional optimization problem and can be hard to characterize. Instead, we can transform the optimization problem to optimize the parameter value given a particular deviation value. Compared to the posterior set \eqref{eq: posterior set} in the $LATE$ example, the general posterior set \eqref{eq: general posterior set} may contains multiple conditional marginal distribution of deviation $\pi_{m_j|F}$. We first discuss the case where $\Pi_{m_j|F}$ is a singleton and then think about more general cases. 

\begin{prop}\label{prop: characterization of the theta_star}
For a given deviation value $m_j(s)=m$, and the observed data distribution $F$, define the intermediate quantity 
	\[
	\begin{split}
	\bar{\theta}(F,m)=\sup_{s: m_j(s)=m,M^s(G^s)=F} \theta(s),\\
	\underline{\theta}(F,m)=\inf_{s: m_j(s)=m,M^s(G^s)=F} \theta(s).
	\end{split}
	\]
	Suppose $\Pi_{m_j|F}$ conatins a singleton $\pi_{m_j|F}^*$, $-\infty<\theta_*\le \theta^*<\infty$, and $\bar{\theta}(F,m), \underline{\theta}(F,m)$ are integrable with respect to $\pi^*_{m_j|F}\pi_{F|\bm{X}_n}$, then the $\theta^*$ and $\theta_*$ can be characterized by 
	\[
	\begin{split}
	\theta^*= \int_{\mathcal{F}}\int_{\mathbb{R}^+}\bar{\theta}(F,m) d\pi^*_{m_j|F} d\pi_{F|\bm{X}_n},\\
	\theta_*= \int_{\mathcal{F}}\int_{\mathbb{R}^+} \underline{\theta}(F,m) d\pi^*_{m_j|F} d\pi_{F|\bm{X}_n}.\\
	\end{split}
	\]
\end{prop}
Proposition \ref{prop: characterization of the theta_star} breaks the optimization problem into model analysis part ($	\bar{\theta}(F,m),\underline{\theta}(F,m)$) and the integration problem. The integration problem can be solved by simulating $(m,F)$ from the product distribution $\pi_{m_j,s|F}^*\times\pi_{F|\bm{X}_n}$. The model analysis part does not have a general solution and is model specific. In the $LATE$ example and subsequently in Section 4, we show that even for nonparametric models, characterizing $\bar{\theta}(F,m)$ and $\underline{\theta}(F,m)$ is feasible. The feasibility comes from the fact that we are not giving up $A_j$ completely: The assumption $A_j$ is initially imposed to facilitate identification of $\theta$ and often a closed-form identified expression is available. When we deviate from $A_j$ by $m_j(s)=m$, such a deviation often results in a tractable change of the econometric structure from the closed-form $A_j$-identified quantity and it is possible to reduce the nonparametric optimization problem to a finite dimensional optimization problem.\footnote{For the $LATE$ application, this is shown in Theorem \ref{thm: Transform infinite optimization to finite optimization}. For an application to the discrete choice model that relaxes the Logit error, see Proposition \ref{prop: discrete choice model relaxation}.} 
\begin{prop} \label{prop: valid confidence set}
	Suppose the conditions in Proposition \ref{prop: characterization of the theta_star} hold. Let $C_{\alpha/2}$ be the lower $\alpha/2$ quantile of the distribution of $\underline{\theta}(F,m)$, and let $C_{1-\alpha/2}$ be the upper $\alpha/2$ quantile of the distribution of $\bar{\theta}(F,m)$, with respect to the product probability $\pi_{m_j|F}\times \pi_{F|\bm{X}_n}$. Then $[C_{\alpha/2},C_{1-\alpha/2}]$ is a valid confidence interval. 
\end{prop}

We now consider the case when $\Pi_{m_j|F}$ contains mutiple prior beliefs. If $\Pi_{m_j|F}$ contains finite many beliefs, then we can simply repeat Proposition \ref{prop: characterization of the theta_star} finitely times and take the union bound. Note that we only need to calculate $\bar{\theta}(F,m)$ and $\underline{\theta}(F,m)$ once and they can be applied for different conditional marginal distribution of deviation. In the following proposition, we show that it is possible to characterize the bounds for a convex prior set.
\begin{prop}\label{prop: convex prior set characterization}
	Suppose $\Pi_{m_j|F}^s$ is a convex set, then in Proposition \ref{prop: characterization of the theta_star}, we just need to focus on the extreme points of $\Pi_{m_j|F}^s$, i.e., we have:
	\[\theta^*=\sup_{\pi_{\theta|\bm{X}_n}\in extrm\left(\Pi_{\theta|\bm{X}_n}\right)}E_{\pi_{\theta|\bm{X}_n}}[\theta],\quad  \theta_*=\inf_{\pi_{\theta|\bm{X}_n}\in extrm\left(\Pi_{\theta|\bm{X}_n}\right)}E_{\pi_{\theta|\bm{X}_n}}[\theta],\]
	where $extrm(\Pi_{\theta|\bm{X}_n})$ is the set of extreme points of $\Pi_{\theta|\bm{X}_n}$.
\end{prop}

Proposition \ref{prop: convex prior set characterization} provides a way to compute $\theta^*$ and $\theta_*$ for many complex prior sets. For example, the set of decreasing conditional marginal density of deviation $\Pi_{m_j|F}^1$ in \eqref{eq: some example of prior sets} has the extreme point sets characterized by step functions.\footnote{Consider the set of weakly decreasing functions defined on $[a,b]$ whose integral is 1, and equip this set with the $||\cdot||_\infty$ norm. Then the extreme points of this set are step functions $f(x)=\mathbbm{1}(a\le x\le a+c)/c$. }

\subsection{Equivalence to Frequentists' Approaches}
We now extend the propositions in Section \ref{section: frequentist Equivalence } to the general framework. We start with the general nonparametric Bayesian convergence assumption. Equip $\mathcal{F}$ with a metric $d_\mathcal{F}$. Let $F_0$ be the true data distribution such that $X_i\sim F_0$, and let $\delta_{F_0}$ be the degenerate distribution with point mass at $F_0$. We use the following Bayesian posterior convergence definition, see Proposition 6.2 in \cite{ghosal2017fundamentals}. Throughout this section, we maintain that the true econometric structure $s_0$ (in the frequentists' framework) satisfies $s_0\in\cap_{l\ne j} A_{l}$.
\begin{assumption}\label{assumption: General convergence Result}
	The posterior distribution $\pi_{F|\bm{X}_n}$ converges weakly to the $\delta_{F_0}$ along the sequence of $\mathbb{P}_{F_0}^{(n)}$, where the $\mathbb{P}_{F_0}^{(n)}$ is the sampling probability measure.\footnote{It is the probability measure of the $\{X_i\}_{i=1}^n$.} 
\end{assumption}
Assumption \ref{assumption: General convergence Result} is the high-level convergence assumption on the posterior. Recall that the definition weak convergence depends on the topology of the sample space $\mathcal{F}$, and hence $d_\mathcal{F}$ matters: If we only care about the CDF, then $d_\mathcal{F}$ can be the Kolmogorov-Smirnov distance; If we care about the density estimation, then $d_\mathcal{F}$ can be chosen as the total variation distance (or equivalently the $L_1$-distance of the density of $F$).

\subsubsection*{Equivalence to the minimal deviation method.}
We start with the definition of the minimal deviation method and the prior set that is equivalent to this minimal deviation method. 
\begin{definition}\label{def: minimal deviation prior}
	Let $m_{j,min}(F)=\min_{s: F=M^s(G^s)} m_j(s)$ be the minimal amount of deviations that is required to rationalize the data distribution $F$. We say a prior $\pi_s$ on $\cap_{l\ne j} A_{l}$ satisfies the minimal deviation constraints if the induced conditional marginal distribution $\pi_{m_j|F}$ has a point mass on $m_{j,min}(F)$ for all $F$.
\end{definition}

\begin{assumption}\label{assump: convergence of parameter integration}
	The bounds $\bar{\theta}(F,m_{j,min}(F))$ and  $\underline{\theta}(F,m_{j,min}(F))$ are continuous at $F_0$. 
\end{assumption} 
Assumption \ref{assump: convergence of parameter integration} is a high-level assumption but it should be easy to pin down by more fundamental conditions. For example, if we can show that $	\bar{\theta}(F,m)$ is a continuous function, and $m_{j,min}(F)$ is a continuous function, then Assumption \ref{assump: convergence of parameter integration} follows for $\bar{\theta}(F,m_{j,min}(F))$.

\begin{prop}\label{prop: minimal deviation equivalence}
	Let the true data distribution be $F_0$ and the conditions in Proposition \ref{prop: characterization of the theta_star} hold. Define the minimal-deviation identified set $\Theta^{ID,min}(F_0)=\{\theta(s): F_0\in M^s(G^s), m_j(s)=m_{j,min}(F_0) \}$.  Suppose the identified set $\Theta^{ID,min}(F_0)$ is bounded, then for any prior $\pi_s$ that satisfies the minimal deviation constraint  and Assumption \ref{assump: convergence of parameter integration}, we can find some large value $M>0$ such that 
	\[
	d_H([\max\{\theta_*,-M\},\min\{\theta^*,M\}], conv\left(\Theta^{ID,min}(F_0)\right)) \rightarrow_p 0,
	\]
	where $d_H$ is the Hausdorff metric, $conv\left(\Theta^{ID,min}(F)\right)$ is the convex hull of the identified set, and $\rightarrow_p$ is convergence in probability with respect to the frequentist sampling probability $\mathbb{P}_{F_0}^{(n)}$. 
\end{prop}
The trimming number $M$ in Proposition \ref{prop: minimal deviation equivalence} helps to avoid irregular tail behaviors in the nonparametric posterior set $\Pi_{\theta|\bm{X}_n}$. The intuition of Proposition \ref{prop: minimal deviation equivalence} is simple: If we put a point mass prior at the minimal deviation amount that rationalize the data, there is no variation in the deviation amount and it is equivalent to let the data to select the minimal deviation amount $(m_{j,min}(F_0))$ and identify $\theta$. However, there are two additional features in Proposition \ref{prop: minimal deviation equivalence}. First, Assumption \ref{assump: convergence of parameter integration} requires the bounds to be continuous as a function of $F$. Intuitively, if the bounds are not continuous, then the weak convergence of the posterior distribution to $F_0$ does not imply the convergence in mean.\footnote{The continuity condition is also crucial for frequentists' estimator to be consistent.} Second, the robust Bayesian posterior bound converges to the convex hull of the frequentists' identified set, but the convexification cannot be avoided because the posterior expectation can be viewed as the Aumann expectation of the identified set \citep{giacomini2021robust}.

\subsubsection*{Equivalence to giving up $A_j$}
Another frequentists' approach is to completely give up $A_j$ and identify the parameter of interest under $\cap_{l\ne j} A_{l}$. We show that the frequentist's identified set under $\cap_{l\ne j} A_{l}$ can also be rationalized by a robust prior set.

\begin{definition}
	Let $m_{j,max}(F)=\min_{s: F=M^s(G^s)} m_j(s)$ be the maximal amount of deviation that can be used to rationalize the data distribution $F$. We say a prior set  $\Pi_{m_j|F}$ satisfies the giving up $A_j$ constraint if it is the class of all distributions supported on $[m_{j,min}(F),m_{j,max}(F)]$. 
\end{definition}
In the definition of $\Pi_{m_j|F}$ above, we simply give up all restrictions on the conditional marginal distribution of deviation except the support condition that is required to ensure the prior set $\Pi_{m_j|F}$ is proper.\footnote{By definition, it is impossible to find a prior belief $\pi_s$ supported on $\cap_{l\ne j} A_{l}$ whose conditional distribution $\pi_{m_j|F}$ with a support wider than $[m_{j,min}(F),m_{j,max}(F)]$. }

\begin{assumption}
	The support bounds $m_{j,min}(F)$ and $m_{j,max}(F)$ are continuous in $F$. Moreover, the function $\theta^*$. Moreover, $	\bar{\theta}(F,m)$ and $\underline{\theta}(F,m)$  are continuous in $(m,F)$. 
\end{assumption}

\begin{prop}\label{assumption: equivalence to Giving up A_j}
	Let $\Theta^{ID}_{-j}(F)=\{\theta(s): s\in \cap_{l\ne j} A_{l}, F=M^s(G^s)\}$  be the frequentist's identified set under $\cap_{l\ne j} A_{l}$. Let $\Pi_{m_j|F}$ be the prior set that satisfies the giving up $A_j$ constraint. If $ \sup \Theta^{ID}_{-j}(F)$ and $ \inf \Theta^{ID}_{-j}(F)$ are continuous in $F$, then there exists a large value of $M>0$ such that 
	\[
	d_H([\max\{\theta_*,-M\},\min\{\theta^*,M\}], conv\left(\Theta^{ID}_{-j}(F_0)\right)) \rightarrow_p 0.
	\]
\end{prop}
In addition to the continuous bounds assumption, we also require the support bounds to be continuous in $F$. This is because we want to ensure the prior set does not change drastically when we change $F$ slightly.

\section{Additional Applications}\label{sec: Additional Applications}
We now examine an application to intersection bounds models and an application to discrete choice models with Logit errors. We illustrate the usefulness of the robust Bayesian method in dealing with refutable models. Appendix \ref{sec: monotone IV + choice of m_j} uses the application to monotone IV models \citep{manski1998monotone} to illustrate subtle issues in choosing the deviation metric $m_j$.

In view of Proposition \ref{prop: characterization of the theta_star}, we aim to show that $\bar{\theta}(F,m)$ and $\underline{\theta}(F,m)$ can be characterized by closed-form expression or by feasible computational method.

\subsection{Intersection Bounds and Moment Inequality Models}
We start with a simple intersection bounds model \citep{chernozhukov2013intersection}. Suppose we have access to observed variables including bound variables $(\bar{Y}_i,\underline{Y}_i)$ and an instrument $Z_i\in [z_l,z_u]$. The observed variables can be rationalized by the following model: $\bar{Y}_i=Y_i+\eta_i^+$, $\underline{Y}_i=Y_i+\eta_i^-$, where the underlying variables $\epsilon_i=(Y_i,\eta_i^+,\eta_i^-,Z_i)$. We are interested in the mean of $Y_i$, that is $\theta(G)=E_{G}[Y_i]$. The rationale behind the intersection bound model is to assume that the instrument $Z_i$ does not the mean of $Y_i$, but it may influence the bounds via $\eta_i^+,\eta_i^-$. 
\begin{assumption}
	The intersection bound assumption $A=A_1\cap A_2$, where $A_1$: $E_G[Y_i|Z_i]=E_G[Y_i]$; and $A_2$: $\inf_{z\in [z_l,z_u]} E_G[\eta_i^+|Z_i=z]\ge0\ge \sup_{z\in [z_l,z_u]} E_G[\eta_i^-|Z_i=z]$.
\end{assumption}
Under the intersection bound assumption, we can derive that, for any $F$ and $F=M^s(G^s)$, we must have $\theta(G)\in [\sup_z E[\underline{Y}_i|Z_i=z], \inf_z E[\bar{Y}_i|Z_i=z]]$. The model is refuted if and only if the interval is empty. We focus on relaxing the $A_2$ assumption while maintaining $A_1$. A simple way to measure the deviation from $A_2$ is the following metric:
\[
m_2(G)= \left( \sup_{z\in [z_l,z_u]} E_G[\eta_i^-|Z_i=z]\right)_+  +\left(-\inf_{z\in [z_l,z_u]} E_G[\eta_i^+|Z_i=z]\right)_+,
\]
where $(t)_+=\max\{0,t\}$. We derive the bounds of $\theta(G)$ in the view of Proposition \ref{prop: characterization of the theta_star}. 
\begin{prop}\label{prop: bounds for intersection bounds model}
	For any observed data distribution $F$, the minimal deviation required to rationalize $F$ is given by $m^{min}_2(F)\equiv \left(\sup_{z\in [z_l,z_u]} E_F[\underline{Y}_i|Z_i=z]-\inf_{z\in [z_l,z_u]} E_F[\bar{Y}_i|Z_i=z]\right)_+$. For any $m\ge m^{min}_2(F)$, we have 
	\[
	\begin{split}
	&\underline{\theta}(F,m)= \sup_{z\in [z_l,z_u]} E_F[\underline{Y}_i|Z_i=z]-m,\\
	&\bar{\theta}(F,m)= \inf_{z\in [z_l,z_u]} E_F[\overline{Y}_i|Z_i=z]+m.
	\end{split}
	\]	
\end{prop}

The intersection bounds model implies a moment inequality model where we can write the model as $E[\bar{Y}_i-\theta|Z_i]\ge 0$ and $E[\theta-\underline{Y}_i|Z_i]\ge 0$. A `reduced-form' way to relax the assumption is to assume  $E[\bar{Y}_i-\theta|Z_i]\ge -m$ while keeping $E[\theta-\underline{Y}_i|Z_i]\ge 0$. In this case, we can characterize $	\bar{\theta}(F,m)= \inf_z E_F[\bar{Y}_i|Z_i=z]+m$ and $\underline{\theta}(F,m)= \sup_z E[\underline{Y}_i|Z_i=z]$. Though relaxing the moment inequality model results in the same bounds on $\theta(G)$ for some choice of deviation metric,\footnote{ Relaxing $E[\bar{Y}_i-\theta|Z_i]\ge -m$ is equivalent to consider a deviation metric $m_j(G)=\inf_{z} \left[E[-\eta_i^+|Z_i=z]\right]_+$.}  we do not recommend directly relaxing the moment inequality bound: Recall that in Section \ref{section: General Theory}, we define $m_j$ as a function on structures rather than on the observed distribution $F$, because we want to ensure the economic interpretation of the deviation value $m$. In contrast, a distance metric defined on $\mathcal{F}$ may not have a direct interpretation from the econometric structures.

\subsection{Discrete Choice and Relaxing Distributional Assumptions}
We now apply our method to study the distributional relaxation in \cite{christensen2023counterfactual}. We consider the discrete choice model where we observe: (1).$Y_i\in \{0,1,2,...,J\}$, the discrete choice outcome; and (2).$Z_i=\{0,1\}$, an indicator of whether choice $J$ is available for individual $i$, and $Z_i=1$ means choice $J$ is available for individual $i$. The standard discrete choice model assumes individual $i$ has a random utility of choosing $j$: $U_{ij}=u_j +\xi_{ij}$, where $\bm{u}\equiv(u_1,...,u_J)$ is the mean utility that does not change across individuals. Choice $j=0$ is the outside option whose utility is normalized to zero.  Consumes choose the object that maximize her utility.

In this example, observed variables are $X_i=(Y_i,Z_i)$ whose distribution is $F$, and underlying variables are $(\xi_{ij},Z_i)$ whose distribution is $G$. A structure $s$ consists of $s=(\bm{u}^s,G^s)$.\footnote{The mapping $M^s$ is determined by $Y_i=\arg\max_j {u_j+\xi_{ij}}$, and $M^s$ can be summarized by the mean utility vector $\bm{u}$. } We are interested in the mean utility parameter for the last choice ($\theta(s)=u_J^s$). We follow \cite{christensen2023counterfactual} to impose the following moment conditions for econometric structure $s$:
\begin{equation}\label{eq: discrete choice moment condition}
\begin{split}
E_{G^s}\left[\mathbbm{1}(U^s_{ij}=\max_{j'=1,...,J}) U^s_{ij'}|Z_i=1\right]&=P_{j|z=1},\\
E_{G^s}\left[\mathbbm{1}(U^s_{ij}=\max_{j'=1,...,J-1}) U^s_{ij'}|Z_i=1\right]&=P_{j|z=0}.\\
\end{split}
\end{equation}
The $P_{j|z}$ is an additional nuisance parameter introduced in \cite{christensen2023counterfactual}, which is identified as $E_F[\mathbbm{1}(Y_i=j)|Z_i=z]$ from the data distribution $F$. In the robust Bayesian framework, $P_{j|z}$ will be estimated by Bayesian methods. 
 
An empirically convenient assumption is to assume that the random utility shocks $\xi_{ij}$ are i.i.d drawn from the Type-I extreme value, which delivers a closed-form expression of the choice probability. However, the Type-I extreme value assumption also induces the I.I.A property on the data identified $P_{j|z}$. Since we have the variation in choice set due to the variation in $Z_i$, the model can be refuted if the I.I.A property fails. 

\cite{christensen2023counterfactual} propose to relax the Type-I extreme value by consider the set of distributions:
\[
\mathcal{G}_m=\left\{G: D(G||G_0)\le m\right\},
\]
where $G_0$ is the Type-I extreme value distribution, $D(G||G_0)$ is the chi-squared divergence between $G$ and $G_0$.\footnote{The chi-squared divergence $D(G||G_0)=\frac{1}{2}\left[\int \frac{(dG)^2}{d(G_0)}d\mu -1\right]$, where $dG$ is the Radon-Nikodym derivatives of $G$ with respect to the dominating measure $\mu$. We do not use the Kullback-Leibler divergence because it is not well defined for Type-I extreme value distribution. See Section 6 of \cite{christensen2023counterfactual} for a discussion.} Our goal is to characterize the parameter bound at a particular Chi-squared divergence level $m$, i.e., we focus on the boundary set $\partial \mathcal{G}_m=\left\{G: D(G||G_0)= m\right\}$ and we want to characterize the bounds:
\begin{equation}\label{eq: discrete choice bounds on mean utility}
\begin{split}
\underline{\theta}(F,m)=\inf_{s: G^s\in \partial \mathcal{G}_m, } u^s_J\quad s.t.\quad (\ref{eq: discrete choice moment condition})\text{ holds},\\
\bar{\theta}(F,m)=\sup_{s: G^s\in \partial \mathcal{G}_m, } u^s_J\quad s.t.\quad (\ref{eq: discrete choice moment condition})\text{ holds}.
\end{split}
\end{equation}
Optimization problems (\ref{eq: discrete choice bounds on mean utility}) are infinite dimensional, but we can use the duality in \cite{christensen2023counterfactual} to recast the problem into a feasible finite dimensional optimization.

\begin{prop}\label{prop: discrete choice model relaxation}
	Solving optimization problems (\ref{eq: discrete choice bounds on mean utility}) are equivalent to solving the
	\begin{equation}\label{eq: discrete choice bounds on mean utility, alternative}
	\begin{split}
	\underline{\theta}(F,m)=\inf u^s_J\quad s.t.\quad m\in (\underline{\Delta}(\bm{u}^s,\bm{P}),\bar{\Delta}(\bm{u}^s,\bm{P})),\\
	\bar{\theta}(F,m)=\sup u^s_J\quad s.t.\quad m\in (\underline{\Delta}(\bm{u}^s,\bm{P}),\bar{\Delta}(\bm{u}^s,\bm{P})).
	\end{split}
	\end{equation}
	where $\bm{P}$ is the vector corresponds to the $P_{j|z}$ in (\ref{eq: discrete choice moment condition}) and depends on $F$ implicitly, and 
	\begin{equation}\label{eq: minimal and maximal chi-squared deviation}
	\begin{split}
	\underline{\Delta}(\bm{u}^s,\bm{P})&=\sup_{\zeta\in \mathbb{R},\lambda\in \mathbb{R}^{2J-1}} -E_{G_0}\left[\phi^\star\left(-\zeta-\lambda'\bm{g}(\bm{u}^s,\bm{\xi})\right)\right]-\zeta-\lambda'\bm{P} \quad s.t.\quad (\ref{eq: discrete choice moment condition})\text{ holds},\\
	\bar{\Delta}(\bm{u}^s,\bm{P})&=\inf_{\zeta\in \mathbb{R},\lambda\in \mathbb{R}^{2J-1}} E_{G_0}\left[\phi^\star\left(\zeta+\lambda'\bm{g}(\bm{u}^s,\bm{\xi})\right)\right]+\zeta+\lambda'\bm{P} \quad \quad \,\,\, s.t.\quad (\ref{eq: discrete choice moment condition})\text{ holds},
	\end{split}
	\end{equation}
	where $\phi^\star(x)=0.5x^2+x$ is the dual function for the Chi-squared divergence, $\bm{g}(\bm{u}^s,\bm{\xi})$ is the vectorized moment function in (\ref{eq: discrete choice moment condition}), whose dimension\footnote{Note that the moment equality for $j=0$ is redundant since the choice probabilities sum to 1. Also, we have one less choice for the $z=0$ case.}  is $2J-1$, and $E_{G_0}$ takes the expectation of $\bm{\xi}$ under the standard i.i.d Logit distribution. In addition, if the maximum or minimum can be achieved in (\ref{eq: minimal and maximal chi-squared deviation}), the corresponding open intervals in (\ref{eq: discrete choice bounds on mean utility, alternative}) should be changed to closed intervals.\footnote{For example, if $	\underline{\Delta}(\bm{u}^s,\bm{P})$ can be achieved by some $\zeta,\lambda$ value while $\bar{\Delta}(\bm{u}^s,\bm{P})$ cannot be achieved, then the intervals in (\ref{eq: discrete choice bounds on mean utility, alternative}) should be changed to $[\underline{\Delta}(\bm{u}^s,\bm{P}),\bar{\Delta}(\bm{u}^s,\bm{P}))$. }
\end{prop}
 It's notable that both (\ref{eq: discrete choice bounds on mean utility, alternative}) and (\ref{eq: minimal and maximal chi-squared deviation}) are finite dimensional optimization problems.
Given a fixed $\bm{u}^s$, \cite{christensen2023counterfactual} interpret $\underline{\Delta}(\bm{u}^s,\bm{P})$ as the minimal Chi-squared divergence relaxation of the $G_0$ that is required to rationalize $\bm{u}^s$ with moment condition (\ref{eq: discrete choice moment condition}), and $\bar{\Delta}(\bm{u}^s,\bm{P})$ is the corresponding maximal Chi-squared divergence relaxation. The optimization \eqref{eq: discrete choice bounds on mean utility, alternative} says that as long as the deviation $m$ is between the minimal and maximal deviation for $\bm{u}^s$, then $\bm{u}^s$ can be rationalized by some underlying distribution $G^s$ with $G^s\in \partial \mathcal{G}$. We then subsequently optimize over $u_J$ to find the bound.  Proposition \ref{prop: discrete choice model relaxation} is not a trivial corollary of \cite{christensen2023counterfactual}, and \eqref{eq: discrete choice bounds on mean utility, alternative} is the additional optimization that we need to compute in the robust Bayesian framework.

\section{Conclusion}\label{section: Conclusion}
This paper proposes a robust Bayesian approach to deal with econometric models with refutable assumptions. The robust Bayesian approach considers a set of prior beliefs that share the same conditional marginal distribution of deviation from the refutable assumption. We propose combining a simulation method and model analysis methods to compute the posterior mean set and the confidence set for the parameter of interest.

As leading applications of the robust Bayesian methods, we study the $LATE$ model, the intersection bounds model, and the discrete choice model with Logit errors. In all three models, it is possible to reduce the infinite-dimensional problems of model analysis to finite-dimensional computational problems. The nice property is likely because of the closed-form characterization of the identified parameter under the refutable models. 

There are several problems that we do not discuss and leave for possible future work. First, we focus on complete models where $M^s$ is a function. In many economic models, such as discrete game models, multiple outcomes can arise, which may require us to define $M^s$ as a multi-valued correspondence. Second, it is interesting to think about the Berstein-von Mise results for the parameter of interests. Note that even if we start with an infinite dimensional model, the parameter of interest is a finite-dimensional parameter, whose asymptotic property may be derived with stronger statistical assumptions. Third, extension to multi-dimensional parameter of interests can be done at the cost of more complicated computation of the $\bar{\theta}(F,m)$ and $\underline{\theta}(F,m)$.

\appendix

\section{Proofs for Results in Section \ref{section: $LATE$application}}
The proof of Lemma \ref{lem: testable implication of IA assumption} is in \cite{kitagawa2015}, and the proof of Lemma \ref{lem: alternative representation of IA-M assumption} is in \cite{liao2020estimating}. Proposition \ref{lem: separation of s|F and F} and Theorem \ref{thm: Transform infinite optimization to finite optimization} are special cases of Proposition \ref{prop: characterization of the theta_star}. Proposition \ref{prop: confidence set consistent, LATE}  is a special case of Proposition \ref{prop: valid confidence set}, and we only check the conditions in Proposition \ref{prop: valid confidence set}. Proofs of Lemmas used in this section can be found in Online Appendix \ref{sec: additional proofs in online appendix}.

\subsection{Proof of Theorem \ref{thm: Transform infinite optimization to finite optimization}}
\begin{proof}
	We only prove the result for $\overline{LATE}(F,m)$ and the result for $\underline{LATE}(F,m)$ holds similarly.  For any $G^s$, we let $g^s(y_1,y_0,d_1,d_0|z)$ denote the density of $G^s$ evaluated at $Y_i(1)=y_1$, $Y_i(0)=y_0$, $D_i(1)=d_1$, $D_i(0)=d_0$ conditional on $Z_i=z$. 	For compliers group and $Z_i=1$, we use the conditional expectation expression to make the following decomposition
	\[
	\begin{split}
		g^s(y_1,y_0,1,0|1)&= \underbrace{ g^s(Y_i(1)=y_1, D_i(1)=1,D_i(0)=0|Z_i=1)}_{\equiv h_1^s(y_1)}\\&\times{ Pr(Y_i(0)=y_0|Y_i(1)=y_1, D_i(1)=1,D_i(0)=0,Z_i=1)}.
	\end{split}
	\]
	Similarly, for compliers group and $Z_i=0$, we have the decomposition:
	\[
	\begin{split}
		g^s(y_1,y_0,1,0|0)&= \underbrace{ g^s(Y_i(0)=y_0, D_i(1)=1,D_i(0)=0|Z_i=0)}_{\equiv h_0^s(y_0)}\\
		&\times{ Pr(Y_i(1)=y_1|Y_i(0)=y_0, D_i(1)=1,D_i(0)=0,Z_i=0)}.
	\end{split}
	\]
	Using the definition of $\overline{LATE}(F,m)$ in \eqref{eq: $LATE$ bound fixing F and m}, we can rewrite $\overline{LATE}(F,m)$ as
	\[
	\begin{split}
		\overline{LATE}(F,m) &= \sup_{s: m^{df}(G^s)=m, M(G^s)=F} 	\frac{\int_{\mathcal{Y}} yh^s_1(y)dy}{\int_{\mathcal{Y}} h^s_1(y)dy}-\frac{\int_{\mathcal{Y}} yh^s_0(y)dy}{\int_{\mathcal{Y}} h^s_0(y)dy}.
	\end{split}
	\]
	
	We first prove two lemmas that equivalently characterize the optimization constraint sets  in \eqref{eq: $LATE$ bound fixing F and m} using the densities $h^s_1$, $h^s_0$ instead of $m^{df}(G^s)=m, M(G^s)=F$. 
	
	\begin{lem}\label{lem: characterizing M(G)=f}
		For any $G^s$ satisfying $M(G^s)=F$ and $s\in A^{TI}\cap A^{EM-NTAT}$, the conditional densities $g^s(y_1,y_0,d_1,d_0|z)$ must satisfy:
		\begin{equation}\label{eq: append, equiv char of M(G^s)=F}
			\begin{split}
				&g^s(y_1,1,1|1)+h^s_1(y_1)=p_F(y_1,1),\\
				&g^s(y_1,0,1|0)=q_F(y_1,1)-g^s(y_1,1,1|0)=q_F(y_1,1)-p_F(y_1,1)+h^s_1(y_1),\\
				&g^s(y_0,0,0|0)+h_0^s(y_0)=q_F(y_0,0),\\
				&g^s(y_0,0,1|1)=p_F(y_0,0)-g^s(y_0,0,0|1)= p_F(y_0,0)-q_F(y_0,0)+h^s_0(y_0),\\
			\end{split}
		\end{equation}
		where $g^s(y_{d_z},d_1,d_0|z)$ marginalizes the density of $g^s(y_1,y_0,d_1,d_0|z)$. 
		Moreover, for any $g^s(y_1,y_0,d_1,d_0|z)$ satisfing \eqref{eq: append, equiv char of M(G^s)=F}, we can  construct an $s\in A^{TI}\cap A^{EM-NTAT}$ such that $M(G^s)=F$. Therefore, \eqref{eq: append, equiv char of M(G^s)=F} is an equivalent characterization of the constraint that $M(G^s)=F$ under $s\in A^{TI}\cap A^{EM-NTAT}$.
	\end{lem}
		
	\begin{lem}\label{lem: characterizing m^df(G^s)=m}
		For any $G^s$ such that $M(G^s)=F$ and $s\in A^{TI}\cap A^{EM-NTAT}$, $m^{df}(G^s)=m$ is equivalent to 
		\[
		Pr_{F}(Z_i=1)\int_{\mathcal{Y}} [q_F(y,1)-p_F(y,1)+h_1^s(y)]dy+ Pr_{F}(Z_i=0)\int_{\mathcal{Y}} [p_F(y,0)-q_F(y,0)+h_0^s(y)]dy=m.
		\]
	\end{lem}	
	
	Using Lemma \ref{lem: characterizing M(G)=f} and Lemma \ref{lem: characterizing m^df(G^s)=m}, we transform the constraint set to constraints on densities:
	\begin{equation}\label{eq: append, optimization using density form}
	\begin{split}
		\overline{LATE}(F,m) &= \sup_{h_1^s,h_0^s} 	\frac{\int_{\mathcal{Y}} yh^s_1(y)dy}{\int_{\mathcal{Y}} h^s_1(y)dy}-\frac{\int_{\mathcal{Y}} yh^s_0(y)dy}{\int_{\mathcal{Y}} h^s_0(y)dy}\\
		\textit{subject to}\quad&\\
		&h_1^s(y)\ge 0,\quad h_0^s(y)\ge 0,\\
		&p_F(y,1)-h_1^s(y)\ge0,\quad q_F(y,0)-h_0^s(y)\ge 0,\\
		& q_F(y,1)-p_F(y,1)+h^s_1(y)\ge 0,\\
		& p_F(y,0)-q_F(y,0)+h_0^s(y)\ge 0,\\
		&Pr_{F}(Z_i=1)\int_{\mathcal{Y}} [q_F(y,1)-p_F(y,1)+h_1^s(y)]dy\\
		+ &Pr_{F}(Z_i=0)\int_{\mathcal{Y}} [p_F(y,0)-q_F(y,0)+h_0^s(y)]dy=m.
	\end{split}
	\end{equation}

	\textit{\textbf{Step 2. Convert the optimization \eqref{eq: append, optimization using density form} to a two-step optimization.}}
	
	\noindent We first note that the optimization \eqref{eq: append, optimization using density form}  can be rewritten as  
	\[
	\begin{split}
	&\quad \quad \quad \max_{h^s_1} \frac{\int_{\mathcal{Y}} yh^s_1(y)dy}{\int_{\mathcal{Y}} h^s_1(y)dy}- \min_{h^s_0}\frac{\int_{\mathcal{Y}} yh^s_0(y)dy}{\int_{\mathcal{Y}} h^s_0(y)dy}\\
	&\quad \quad \quad \textit{subject to}\\
	&\quad \quad \quad\max\{p_F(y,1)-q_F(y,1),0\}\le h^s_1(y)\le p_F(y,1),\\
	&\quad \quad \quad\max\{q_F(y,0)-p_F(y,0),0\}\le h^s_0(y)\le q_F(y,0),\\
	&\int_{\mathcal{Y}}{q_F(y,1)-p_F(y,1)+h^s_1(y)}dy Pr(Z_i=1) +  \int_{\mathcal{Y}}{p_F(y,0)-q_F(y,0)+h^s_0(y)}dy Pr(Z_i=0)=m,
	\end{split}
	\]
	where we separate the optimization for $h_1^s$ and $h_0^s$, and simplify the constraint set. 
	
	This is still an infinite-dimensional optimization problem. However, we can further transform it into a two-step optimization by constraining the value of integration value of $h^s_1$ and $h^s_0$. By setting the value of $\int_{\mathcal{Y}} h^s_1(y) dy=a$ and $\int_{\mathcal{Y}} h^s_0(y) dy=b$, we can further transform the optimization as
	\begin{equation}\label{eq: append, two layer optimization}
	\begin{split}
		&\max_{a,b}\left\{\max_{h^s_1} \frac{\int_{\mathcal{Y}} yh^s_1(y)dy}{a}- \min_{h^s_0}\frac{\int_{\mathcal{Y}} yh^s_0(y)dy}{b}\right\}\\
		&\max\{p_F(y,1)-q_F(y,1),0\}\le h^s_1(y)\le p_F(y,1),\\
		&\max\{q_F(y,0)-p_F(y,0),0\}\le h^s_0(y)\le q_F(y,0),\\
		&\int_{\mathcal{Y}} h^s_1(y) dy=a,\quad \int_{\mathcal{Y}} h^s_0(y) dy=b,\\
		&\quad aPr_F(Z_i=0)+Pr_F(D_i=1)-Pr_F(D_i=1|Z_i=1)Pr_F(Z_i=0)\\
		&+bPr_F(Z_i=1)+Pr_F(D_i=0)-Pr_F(D_i=0|Z_i=0)Pr_F(Z_i=1)=m,\\
		&\int_{\mathcal{Y}} \max\{p_F(y,1)-q_F(y,1),0\}dy \le a\le \int_{\mathcal{Y}} p_F(y,1) dy,\\
		&\int_{\mathcal{Y}} \max\{q_F(y,0)-p_F(y,0),0\}dy \le b \le \int_{\mathcal{Y}} q_F(y,0)dy.
	\end{split}
	\end{equation}
	The last two constraints of \eqref{eq: append, two layer optimization} are implied by the first four constraints but we write it down explicitly because it will be used later.
	
	\textit{\textbf{Step 3. The closed-form solution for the inner optimization of \eqref{eq: append, two layer optimization}.}}
	We now derive the closed-form solution for the inner optimization 
	\begin{equation}\label{eq: append, optimization with fixed a and b}
	\begin{split}
		&\max_{h^s_1} \frac{\int_{\mathcal{Y}} yh^s_1(y)dy}{a}- \min_{h^s_0}\frac{\int_{\mathcal{Y}} yh^s_0(y)dy}{b}\\
		&\textit{subject to}\\
		&\max\{p_F(y,1)-q_F(y,1),0\}\le h^s_1(y)\le p_F(y,1),\\
		&\max\{q_F(y,0)-p_F(y,0),0\}\le h^s_0(y)\le q_F(y,0),\\
		&\int_{\mathcal{Y}} h^s_1(y) dy=a,\quad \int_{\mathcal{Y}} h^s_0(y) dy=b.\\
	\end{split}
	\end{equation}
	We can transform the density as $h^s_1(y)= \max\{p_F(y,1)-q_F(y,1),0\} + \tilde{h}_1(y)$ and $h^s_0(y)= \max\{q_F(y,0)-p_F(y,0),0\} + \tilde{h}_0(y)$, where 
	\[
	\begin{split}
		0\le \tilde{h}_1(y) \le \min\{p_F(y,1),q_F(y,1)\},\quad \int_{\mathcal{Y}} \tilde{h}_1(y) dy=  a- \int_{\mathcal{Y}} \max\{p_F(y,1)-q_F(y,1),0\}dy,\\
		0\le \tilde{h}_0(y) \le \min\{p_F(y,0),q_F(y,0)\},\quad \int_{\mathcal{Y}} \tilde{h}_0(y) dy=  b- \int_{\mathcal{Y}} \max\{q_F(y,0)-p_F(y,0),0\}dy.
	\end{split}
	\]
	The $\tilde{h}_1$ and $\tilde{h}_0$ can be viewed as the densities in addition to the minimal density requirement. 
	
	To maximize $\int_\mathcal{Y} y (\max\{p_F(y,1)-q_F(y,1),0\} + \tilde{h}_1(y))dy$ subject to the constraints on $\tilde{h}_1$, we must exhaust all possible large values of $y$ and allocate them to $\tilde{h}_1(y)$. That is, the maximizer $\tilde{h}^*_1(y)$ satisfies 
	\[
	\int_{\mathcal{Y}}\tilde{h}^*_1(y)dy= \min\{p_F(y,1),q_F(y,1)\}\mathbbm{1}(y> \bar{y}_{max})
	\]
	and 
	\[
	\int_{\bar{y}_{max}}^{\infty} \min\{p_F(y,1),q_F(y,1)\} dy= a- \int_{\mathcal{Y}} \max\{p_F(y,1)-q_F(y,1),0\}dy.
	\]
	The maximizer above is the $h^{max}_1$ in Theorem \ref{thm: Transform infinite optimization to finite optimization}. The proof for the optimization for $\tilde{h}_0$ is the same. Combining the closed-form in Step 3 and the two-step optimization result in Step 2, we finish the proof.
	\end{proof}
	
	\subsection{Proof of Proposition \ref{prop: Frequentist to minimal deviation}}
	\begin{proof}
	If we impose the class $\Pi_{s}^{min}$ that satisfies the minimal defier constraints, then for any $F$, we can derive the minimal amount of compliers from Theorem \ref{thm: Transform infinite optimization to finite optimization} 
	\[
	\begin{split}
		a(F)\equiv\int_{\mathcal{Y}}(p_F(y,1)-q_F(y,1)) \mathbbm{1}(p_F(y,1)-q_F(y,1)\ge 0) dy,\\
		b(F)\equiv\int_{\mathcal{Y}}(q_F(y,0)-p_F(y,0)) \mathbbm{1}(q_F(y,0)-p_F(y,0)\ge 0) dy.
	\end{split}
	\]
	where the $a(F)$ (resp. $b(F)$) is the minimal probability of compliers conditional on $Z_i=1$ (resp. $Z_i=0$). Since by Lemma \ref{lem: characterizing M(G)=f}, the density of defiers increases with the density of compliers, minimal probability of compliers implies the minimal probability of defiers. Therefore, the minimal defier constraints implies that $a(F)$ and $b(F)$ are the corresponding conditional probability of compliers.
	
	Since $a(F)$ and $b(F)$ are unique given the minimal defiers, the marginal densities $h_1^s$ and $h_0^s$ achieving the minimal compliers are also uniquely pinned down as 
	\[
	h_1^s= \max\{p_F(y,1)-q_F(y,1),0\},\quad h_0^s= \max\{q_F(y,0)-p_F(y,0),0\},
	\]and $\overline{LATE}(F,m^{df}_{min}(F))=\underline{LATE}(F,m^{df}_{min}(F))$ must hold. Moreover, the $\overline{LATE}(F,m^{df}_{min}(F))$ must satisfy
	\begin{equation}\label{eq: append, $LATE$ bar for minimal deviation}
		\overline{LATE}(F,m^{df}_{min}(F))= \frac{\int_{\mathcal{Y}_1(F)} y(p_{F}(y,1)-q_{F}(y,1)) dy}{\int_{\mathcal{Y}_1(F)} p_{F}(y,1)-q_{F}(y,1)dy} - \frac{\int_{\mathcal{Y}_0(F)} y(q_{F}(y,0)-p_{F}(y,0)) dy}{\int_{\mathcal{Y}_0(F)} q_{F}(y,0)-p_{F}(y,0)dy},
	\end{equation}
	where the integration region $\mathcal{Y}_1(F)$ depends on the $F$ rather than $F_0$.

	In view of Proposition \ref{prop: minimal deviation equivalence}, we must show (\ref{eq: append, $LATE$ bar for minimal deviation}) is continuous in $F$ with respect to the total variation metric on $\mathcal{F}$. We will use the fact that convergence in total variation is equivalent to convergence in $L_1$ distance. As a result, $F\rightarrow F_0$ in $d_\mathcal{F}$ metric implies $p_F(y,d)\rightarrow_{L_1} p_{F_0}(y,d)$ and $q_F(y,d)\rightarrow_{L_1} q_{F_0}(y,d)$ for $d=0,1$. Using the Lemma \ref{lem: convergence of integratino on level set} below, for any $F\rightarrow F_0$ in $L_1$ norm, we must have 
	\[
	\left|\int_{\mathcal{Y}_1(F)} (p_{F}(y,1)-q_{F}(y,1))dy - \int_{\mathcal{Y}_1(F_0)} (p_{F_0}(y,1)-q_{F_0}(y,1))dy\right|\rightarrow 0,
	\]
	and since $|y|$ is bounded by a constant $M_y$, H\"oder inequality implies 
	\[
	\begin{split}
		&\quad \left|\int_{\mathcal{Y}_1(F)} y(p_{F}(y,1)-q_{F}(y,1))dy - \int_{\mathcal{Y}_1(F_0)} y(p_{F_0}(y,1)-q_{F_0}(y,1))dy\right|\\
		&\le  M_y\left|\int_{\mathcal{Y}_1(F)} (p_{F}(y,1)-q_{F}(y,1)dy - \int_{\mathcal{Y}_1(F_0)} (p_{F_0}(y,1)-q_{F_0}(y,1))dy\right|\rightarrow 0.
	\end{split}
	\]
	Similar convergence holds for the numerator and denominator of the second term in \eqref{eq: append, $LATE$ bar for minimal deviation}. We can show the continuity of $\overline{LATE}(F,m^{df}_{min}(F))$ at $F_0$ using composition function. 
\end{proof}

\begin{lem}\label{lem: convergence of integratino on level set}
	For any sequence of functions $f_n$ such that $\int_{-M_y}^{M_y} |f_n(y)-f_0(y)|dy\rightarrow 0$, we must have
	\[\left|\int_{\{y:f_n(y)\ge0\} } f_n(y)dy-\int_{\{y:f_0(y)\ge0\} } f_0(y)dy\right|\rightarrow 0.\]
\end{lem}

\subsection{Proof of Proposition \ref{prop: consistent posterior}}
\begin{proof}
	Proposition \ref{prop: consistent posterior} is standard in the nonparametric Bayesian density estimation. We only direct the readers to the following textbook source: Theorem 7.13 and Example 7.14 in \cite{ghosal2017fundamentals} show that the normal mixture model (\ref{eq: conditional prior for theta and sigma})-(\ref{eq: distribution for ydz}) for density estimation is $L_1$-consistent for $p_{F_0}(y,d)$ and $q_{F_0}(y,d)$ as long as $\alpha>1$ and these four densities are in the Kullback-Leibler support of the prior belief $\text{Dir}(H_{dz})$. 
	
	To check the Kullback-Leibler support condition, we use Theorem 7.3  in \cite{ghosal2017fundamentals}: The (B1) condition is ensured by the normal pdf function, (B2)-(B3) are assumed in Assumption \ref{assumption: consistency requirement for normal kernel}, and (B4) is satisfied by the normal pdf function (as shown in Example 7.5).
	
	The posterior of $\pi_{dz|\bm{X}_n}$ is consistent for the true distribution of $(D_i,Z_i)$ by Theorem 6.16 and Example 6.20. The posterior as the product of the two posterior densities is consistent by continuous mapping theorem. 
\end{proof}
\section{Proofs of Results in Section \ref{section: General Theory}}

\subsection{Proofs of Proposition \ref{prop: characterization of the theta_star}}
\begin{proof}
	We prove the result for $\theta^*$ and the result for $\theta_*$ follows similarly. 
	
	First note that we can write $\theta^*$ as 
	\begin{equation}\label{eq: append, proving theta^* 1}
	\begin{split}
		\theta^*&=\sup_{\pi_{s|\bm{X}_n}\in \Pi_{s|\bm{X}_n}}\int_{\pi_{s|\bm{X}_n}} \theta(s) d\pi_{s|\bm{X}_n}\\
		&= \sup_{\pi_{s|F}:\pi_{m_j|F}=\pi^*_{m_j|F}} \int_{\mathcal{F}} \int_{\mathcal{S}} \theta(s)d\pi_{s|F} d\pi_{F|\bm{X}_n}\\
		&=_{(*)} \int_{\mathcal{F}} \left[\sup_{\pi_{s|F}: \pi_{m_j|F}=\pi^*_{m_j|F}} \int_{\mathcal{S}} \theta(s)d\pi_{s|F}\right] d\pi_{F|\bm{X}_n}.
	\end{split}
	\end{equation}
	where the first equality of the equation above holds by definition and the second equality holds by the definition of the posterior set and the unique conditional marginal distribution $\pi_{m_j|F}^*$. In the last step $(*)$, we change the order of optimization and integration, which is proved in the following. 
	
	\paragraph{Proving Step $(*)$}To see this, since $\theta^*$ is finite, for any $\epsilon>0$ and ($\pi_{F|\bm{X}_n}$-almost surely) any $F$, we can find a $\tilde{\pi}_{s|F}$ whose marginal distribution $\tilde{\pi}_{m_j|F}=\pi^*_{m_j|F}$ such that 
	\[
	\int_{\mathcal{S}}\theta(s) d\tilde{\pi}_{s|F}> \sup_{\pi_{s|F}:\pi_{m_j|F}=\pi^*_{m_j|F}} \int_{\mathcal{S}} \theta(s)d\pi_{s|F}-\epsilon.
	\]
	By constructing a posterior $\tilde{\pi}_{s|\bm{X}_n}=\tilde{\pi}_{s|F}\times \pi_{F|\bm{X}_n}$, we can show that 
	\begin{equation}\label{eq: append, one direction}
		\begin{split}
			\sup_{\pi_{s|F}:\pi_{m_j|F}=\pi^*_{m_j|F}} \int_{\mathcal{F}} \int_{\mathcal{S}} \theta(s)d\pi_{s|F} d\pi_{F|\bm{X}_n}&\ge_{(a)} \int_{\mathcal{F}} \int_{\mathcal{S}} \theta(s)d\tilde{\pi}_{s|F} d\pi_{F|\bm{X}_n} \\ 
			&\ge\int_{\mathcal{F}} \left[\sup_{\pi_{s|F}:\pi_{m_j|F}=\pi^*_{m_j|F}} \int_{\mathcal{S}} \theta(s)d\pi_{s|F}\right] d\pi_{F|\bm{X}_n}-\epsilon.
		\end{split}
	\end{equation}
	where step $(a)$ follows by definition and that $\tilde{\pi}_{s|F}\in \Pi_{s|F}$ by construction.  \eqref{eq: append, one direction} proves one direction of the $(*)$ step.
	
	Conversely, since $\theta^*$ is finite, let $\breve{\pi}_{s|\bm{X}_n}$ be the belief such that 
	\begin{equation}\label{eq: append, the other direction, intermediate step}
		\int_{\mathcal{S}} \theta(s) d\breve{\pi}_{s|\bm{X}_n} \ge \sup_{\pi_{s|\bm{X}_n}\in \Pi_{s|\bm{X}_n}}\int_{\mathcal{S}} \theta(s) d\pi_{s|\bm{X}_n}-\epsilon.
	\end{equation}
	The belief $\breve{\pi}_{s|\bm{X}_n}$ can be written as the conditional form $\breve{\pi}_{s|F}\times \pi_{F|\bm{X}_n}$. As a result 
	\begin{equation}\label{eq: append, the other direction, final step}
		\begin{split}
			\int_{\mathcal{F}} \left[\sup_{\pi_{s|F}: \pi_{m_j|F}=\pi^*_{m_j|F}} \int_{\mathcal{S}} \theta(s)d\pi_{s|F}\right]d\pi_{F|\bm{X}_n}&\ge_{(b)} \int_{\mathcal{F}} \left[\int_{\mathcal{S}} \theta(s)d\breve{\pi}_{s|F}\right]d\pi_{F|\bm{X}_n}\\
			&\ge_{(c)} \int_{\mathcal{F}} \left[\sup_{\pi_{s|F}: \pi_{m_j|F}=\pi^*_{m_j|F}} \int_{\mathcal{S}} \theta(s)d\pi_{s|F}\right] d\pi_{F|\bm{X}_n}-\epsilon.
		\end{split}
	\end{equation}
	where $(b)$ follows by definition and $(c)$ follows by \eqref{eq: append, the other direction, intermediate step} and the second equality of \eqref{eq: append, proving theta^* 1}.  Since $\epsilon$ is arbitrary, \eqref{eq: append, the other direction, final step} proves the other direction of  $(*)$.
	
	It remains to show that 
	\begin{equation}\label{eq: append, proving theta^* 2}
	\sup_{\pi_{s|F}: \pi_{m_j|F}=\pi^*_{m_j|F}} \int_{\mathcal{S}} \theta(s)d\pi_{s|F}= \int_{\mathbb{R}^+}\bar{\theta}(F,m) d\pi^*_{m_j|F}.
	\end{equation}
	
	Since $\theta^*<\infty$ implies that $\sup_{\pi_{s|F}: \pi_{m_j|F}=\pi^*_{m_j|F}} \int_{\mathcal{S}} \theta(s)d\pi_{s|F}$ is $\pi_{F|\bm{X}_n}$-almost surely finite, for $\pi_{F|\bm{X}_n}$-a.s. all $F$, we can find a belief $\breve{\pi}_{s|F}$ whose $\breve{\pi}_{m_j|F}=\pi^*_{m_j|F}$ such that 
	\[
	\int_{\mathcal{S}} \theta(s)d\breve{\pi}_{s|F}\ge  \sup_{\pi_{s|F}: \pi_{m_j|F}=\pi^*_{m_j|F}} \int_s
	\theta(s)d\pi_{s|F}-\epsilon.
	\]
	We can further conditional on $m_j(s)=m$ to write 
	\[
	\int_{\mathcal{S}} \theta(s)d\breve{\pi}_{s|F}=\int_m \int_{\mathcal{S}} \theta(s) d\pi_{s|m,F} d\pi_{m_j|F},
	\]
	where $\pi_{s|m,F}$ is the distribution of $s$ conditional on $m_j(s)=m$ and $F=M^s(G^s)$. Since for all $s'$ in the support of $\pi_{s|m,F}$ we have $\theta(s')\le \bar{\theta}(F,m)$, we can conclude that 
	\[
	\begin{split}
		\int_{\mathbb{R}^+}\bar{\theta}(F,m) d\pi_{m_j,s|F}&\ge \int_{\mathcal{S}} \theta(s)d\breve{\pi}_{s|F}\\
		&\ge \sup_{\pi_{s|F}: \pi_{m_j|F}=\pi^*_{m_j|F}} \int_{\mathcal{S}}
		\theta(s)d\pi_{s|F}-\epsilon.
	\end{split} 
	\]
	Since $\epsilon$ is arbitrary, we can conclude that $    \sup_{\pi_{s|F}: \pi_{m_j|F}=\pi^*_{m_j|F}} \int_{\mathcal{S}} \theta(s)d\pi_{s|F} \le \int_{\mathbb{R}^+}\bar{\theta}(F,m) d\pi^*_{m_j|F}$. 
	
	For the converse direct, since $\bar{\theta}(F,m)$ is integrable with respect to $\pi^*_{m_j|F} \times \pi_{F|\bm{X}_n}$, then $\bar{\theta}(F,m)$ is finite almost surely. For any $\epsilon>0$, ($\pi_{F|\bm{X}_n}\times \pi^*_{m_j|F}$-almost surely) any $F$ and any $m$, we can find structure $\tilde{s}_m$ such that 
	\[
	\begin{split}
		\theta(\tilde{s}_m)\ge\bar{\theta}(F,m)-\epsilon,\\
		m_j(\tilde{s}_m)=m,\quad \quad M^{\tilde{s}_m}(G^{\tilde{s}_m})=F.
	\end{split}
	\]
	We construct a conditional prior $\tilde{\pi}_{s|F}$ supported on $\cup_{m} \{\tilde{s}_m\}$ such that $\tilde{\pi}_{s|F}(\cup_{m\in B} \{\tilde{s}_m\})=\pi^*_{m_j|F}(B)$. As a result, $\tilde{\pi}_{s|F}$ is in the feasible set, and 
	\[
	\begin{split}
		\sup_{\pi_{s|F}: \pi_{m_j|F}=\pi^*_{m_j|F}} \int_{\mathcal{S}} \theta(s)d\pi_{s|F}\ge \int_{\mathcal{S}} \theta(s) d\tilde{\pi}_{{s}|F}\ge \int_{\mathbb{R}^+} \bar{\theta}(F,m)d\pi^*_{m_j|F}-\epsilon.
	\end{split}
	\]
	Since $\epsilon$ is arbitrarily small, we prove the other direction of \eqref{eq: append, proving theta^* 2}. The prove is done by combining \eqref{eq: append, proving theta^* 1} and  \eqref{eq: append, proving theta^* 2}. 
\end{proof}

\subsection{Proof of Proposition \ref{prop: valid confidence set} }
\begin{proof}
	It is easy to see that for any structure $s$ such that $M^s(G^s)=F$ and $m_j(s)=m$, $\theta(s)\in[\underline{\theta}(F,m),\bar{\theta}(F,m)]$ must hold by definition. As a result, for any $\pi_{\theta|\bm{X}_n}\in \Pi_{\theta|\bm{X}_n}$:
	\[
	\begin{split}
	Pr_{\pi_{\theta|\bm{X}_n}} (\theta\notin [C_{\alpha},C_{1-\alpha/2}])&=Pr_{\pi_{s|F}\times \pi_{F|\bm{X}_n}}(\theta(s)\notin [C_{\alpha/2},C_{1-\alpha/2}] )\\
	&=Pr_{\pi_{s|F}\times \pi_{F|\bm{X}_n}}(\theta(s)>C_{1-\alpha/2} ) + Pr_{\pi_{s|F}\times \pi_{F|\bm{X}_n}}(\theta(s)<C_{\alpha/2} )\\
	&\le Pr_{\pi^*_{m_j|F}\times \pi_{F|\bm{X}_n}}(\bar{\theta}(F,m)>C_{1-\alpha/2} ) + Pr_{\pi^*_{m_j|F}\times \pi_{F|\bm{X}_n}}(\underline{\theta}(F,m)>C_{\alpha/2} ) \\
	&=\alpha.
	\end{split}
	\]
Finally, take $\sup$ of the above equation over all beliefs in $\Pi_{\theta|\bm{X}_n}$ to finish the proof. 
\end{proof}

\subsection{Proof of Proposition \ref{prop: convex prior set characterization}}
\begin{proof}
	If $\pi^{mix}_{m_j|F}=\alpha\pi^1_{m_j|F}+(1-\alpha) \pi^2_{m_j|F}$, then the linearity of integration implies that $\theta^*|_{\pi^{mix}_{m_j|F}}=\alpha\theta^*|_{\pi^{1}_{m_j|F}}+(1-\alpha)\theta^*|_{\pi^{2}_{m_j|F}}$, where $\theta^*|_{\tilde{\pi}}$ denotes the value of $\theta^*$ when the conditional marginal belief of deviation is $\tilde{\pi}$. So the maximum of $\theta^*$ is achieved at the extreme point. Similar results holds for the lower bound.
\end{proof}

\subsection{Proof of Proposition \ref{prop: minimal deviation equivalence}}

\begin{proof}
	We prove the statement for $\theta^*$ and the result for $\theta_*$ follows similarly. Take $M$ to be any large number such that $M>\sup \Theta^{ID,min}(F_0)$. 	By Assumption \ref{assumption: General convergence Result}, the continuity of $\theta^*$ in Assumption \ref{assump: convergence of parameter integration}, and the definition of weak convergence, we must have $\int_{\mathcal{F}} \min\{\bar{\theta}(F,m_{j,min}(F)),M\} d\pi_{F|\bm{X}_n}\rightarrow  \min\{\bar{\theta}(F_0,m_{j,min}(F_0)),M\}$ along the empirical distribution sequence $\mathbb{P}_{F_0}^{(n)}$. Therefore, 
	\[
	\begin{split}
	\min\{\theta^*,M\}&=_{(i)} \int_{\mathcal{F}}\int_{\mathbb{R}^+} \min\{\bar{\theta}(F,m),M\} d\pi_{m_j,s|F} d\pi_{F|\bm{X}_n}\\
	&=_{(ii)}\int_{\mathcal{F}} \min\{\bar{\theta}(F,m_{j,min}(F)) ,M\}d\pi_{F|\bm{X}_n}\\&\rightarrow \min\{\bar{\theta}(F_0,m_{j,min}(F_0)),M\}= \bar{\theta}(F_0,m_{j,min}(F_0))\\
	\end{split}
	\]
	where $(i)$ follows by slightly adapting the proof of Proposition \ref{prop: characterization of the theta_star}, $(ii)$ follows because $\pi_s$ staifies the minimal deviation constraints.
	
	By the definition of $\bar{\theta}(F,m)$ in Proposition \ref{prop: characterization of the theta_star}, we have $\bar{\theta}(F_0,m_{j,min}(F_0))= \sup \{\theta(s): F_0=M^s(G^s), m_j(s)=m_{j,min}(F_0) \}=\sup \Theta^{ID,min}(F_0)$. The proposition follows. 
\end{proof} 

\subsection{Proof of Proposition \ref{assumption: equivalence to Giving up A_j}}

\begin{proof}
	
	\textbf{We first show that $\lim\inf \min\{\theta^*,M\}\ge  \sup\Theta^{ID}_{-j}(F_0)$. } 
	Take $M>|\sup  \Theta^{ID}_{-j}(F_0)|$. Starting from any $\tilde{\theta}\in \Theta^{ID}_{-j}(F_0)$, we can find an $\tilde{s}$ such that $\theta(\tilde{s})=\tilde{\theta}$ and $F_0= M^{\tilde{s}}(G^{\tilde{s}})$. We define $\tilde{m}= m_j(\tilde{s})$.  For any $F$ in the support of the posterior $\pi_{F|\bm{X}_n}$, let $\breve{m}(F)=\min\{ \max \{ \tilde{m}, m_{j,min}(F) \},m_{j,max}(F)\}$, which is the projection of $\tilde{m}$ onto the interval $[m_{j,min}(F), m_{j,max}(F)]$. 
	
	Consider the degenerate marginal belief $\tilde{\pi}_{m_j|F}=\delta_{\breve{m}(F)} \in \Pi_{m_j|F}$.\footnote{This degenerate prior belief is in the  prior belief set that gives up $A_j$.} Then 

		\begin{equation}\label{eq: append, giving up A_j equivalence 0}
			\min\{\theta^*,M\}\ge_{(a)} \int_{\mathcal{F}} \min\{\bar{\theta}(F,\breve{m}(F)),M\} d\pi_{F|\bm{X}_n}\ge_{(b)} \min\{\tilde{\theta},M\}=\tilde{\theta}
	\end{equation} 
	where $(a)$ follows because by the definition of $\theta^*$ and that $\bar{\theta}(F,\breve{m}(F))=\int \bar{\theta}(F,m) d\tilde{\pi}_{m_j|F}$. Our goal is to show:
	\begin{equation}\label{eq: append, giving up A_j equivalence 1}
		\lim\inf \int_{\mathcal{F}} \min\{\bar{\theta}(F,\breve{m}(F)),M\} d\pi_{F|\bm{X}_n}\ge \min\{\tilde{\theta},M\}=\tilde{\theta}.
	\end{equation} 
	Then by combining \eqref{eq: append, giving up A_j equivalence 0} and \eqref{eq: append, giving up A_j equivalence 1}, we can show $\lim\inf \min\{\theta^*,M\}\ge \tilde{\theta}$ for all $\tilde{\theta}\in \Theta^{ID}_{-j}(F_0)$, and hence 
	\begin{equation}\label{eq: append, giving up A_j equivalence 1.5}
		\lim\inf \min\{\theta^*,M\}\ge  \sup\Theta^{ID}_{-j}(F_0).
	\end{equation}

	By Assumption \ref{assumption: equivalence to Giving up A_j}, the continuity of $m_{j,min}(F)$ and $m_{j,max}(F)$ implies the continuity of $\breve{m}(F)$. The continuity of $\bar{\theta}(F,m)$ and $\underline{\theta}(F,m)$  implies the continuity of $\theta(\breve{m}(F);F) $ in $F$, and hence the function $\min\{\theta(\breve{m}(F);F) ,M\}$. We then apply the definition of weak convergence of $\pi_{F|\bm{X}_n}$ to get 
	\begin{equation}\label{eq: append, giving up A_j equivalence 2}
		\lim \inf \int_{\mathcal{F}} \min\{ \bar{\theta}(F,\breve{m}(F)) ,M\}d\pi_{F|\bm{X}_n} \rightarrow \min\{\bar{\theta}(F_0,\breve{m}(F_0)),M\}.
		\end{equation}
	
	Recall that for the $\tilde{s}$ that generates $\tilde{\theta}$ and $F_0$, we have $m_j(\tilde{s})=\tilde{m}$, then for this $\tilde{s}$, we must have $\tilde{m}\in [m_{j,min}(F_0),m_{j,max}(F_0)]$, and the definition of $\breve{m}(\cdot)$ implies that $\breve{m}(F_0)=\tilde{m}$ must hold.\footnote{Since $\tilde{s}$ generate $F_0$, then $\tilde{m}$ must lie in the interval  $[m_{j,min}(F_0),m_{j,max}(F_0)]$.} Then by the definition of $\bar{\theta}(F,m)$, we have 
	\begin{equation}\label{eq: append, giving up A_j equivalence 3}
	\bar{\theta}(F_0,\breve{m}(F_0))=\sup_{s: m_j(s)=\tilde{m},F_0=M^s(G^s)} \theta(s)\ge \theta(\tilde{s})=\tilde{\theta}.
	\end{equation}
	Combine \eqref{eq: append, giving up A_j equivalence 2} and \eqref{eq: append, giving up A_j equivalence 3} to show inequality \eqref{eq: append, giving up A_j equivalence 1}. 
	
	\textbf{We then show  $[\max\{\theta_*,-M\},\min\{\theta^*,M\}] \subseteq \overline{conv(\Theta^{ID}_{-j}(F_0))}$ with probability converging to 1.}  It suffices to show that, $\min\{\theta^*,M\}\rightarrow_p\sup  \Theta^{ID}_{-j}(F_0)$ and $\max\{\theta_*,-M\}\rightarrow_p\inf  \Theta^{ID}_{-j}(F_0)$.  To see this, following the logic of proofs of Proposition \ref{prop: characterization of the theta_star}, for any $\epsilon>0$, we can find a $\tilde{\pi}_{s|F}$  such that 
	\begin{equation}\label{eq: append, giving up A_j equivalence 4} 
	\int_{\mathcal{F}} \int_{\mathcal{S}} \min\{\theta(s),M\} d\tilde{\pi}_{s|F} d\tilde{\pi}_{F|\bm{X}_n}\ge \min\{\theta^*,M\}-\epsilon.
	\end{equation}
	We can also find a $\tilde{s}$ in the support $\tilde{\pi}_{s|F}$ such that $\theta(\tilde{s})\ge \sup_{s\in supp(\tilde{\pi}_{s|F})}\theta(s)-\epsilon$. Let $\breve{\pi}_{s|F}$ be the degenerate distribution with point mass on $\tilde{s}$, and we must have 
	\begin{equation}\label{eq: append, giving up A_j equivalence 5}
	\int_{\mathcal{F}} \int_{\mathcal{S}} \min\{\theta(s),M\} d\breve{\pi}_{s|F} d\tilde{\pi}_{F|\bm{X}_n}\ge \int_{\mathcal{F}} \int_{\mathcal{S}} \min\{\theta(s)-\epsilon,M\} d\tilde{\pi}_{s|F} d\tilde{\pi}_{F|\bm{X}_n} \ge_{(c)} \min\{\theta^*,M\}-2\epsilon,
	\end{equation}
	where we use \eqref{eq: append, giving up A_j equivalence 4}  to derive $(c)$.
	
	Note that by the generacy of $\breve{\pi}_{s|F}$ and that $F= M^{\tilde{s}}(G^{\tilde{s}})$, we must have 
	\begin{equation}\label {eq: append, giving up A_j equivalence 6}
	\int_{\mathcal{S}} \min\{\theta(s),M\} d\breve{\pi}_{s|F} = \min\{\theta(\tilde{s}),M\}\le  \min\{\sup\Theta^{ID}_{-j}(F),M\}.
	\end{equation} Combining \eqref{eq: append, giving up A_j equivalence 5} and \eqref{eq: append, giving up A_j equivalence 6}, we can derive
	\begin{equation}\label{eq: append, giving up A_j equivalence 7} 
		{\int_{\mathcal{F}}  \min\{\sup\Theta^{ID}_{-j}(F),M\} d\pi_{F|\bm{X}_n}}\ge \int_{\mathcal{F}} \int_{\mathcal{S}} \theta(s) d\breve{\pi}_{s|F} d\tilde{\pi}_{F|\bm{X}_n}\ge \theta^*-2\epsilon.
	\end{equation} 
	Take $\lim\sup$ of both sides of \eqref{eq: append, giving up A_j equivalence 7}, the LHS converges to $ \min\{\sup\Theta^{ID}_{-j}(F_0),M\}$ by the weak convergence of $\tilde{\pi}_{F|\bm{X}_n}$ and the bounded continuity of $\min\{\sup\Theta^{ID}_{-j}(F),M\}$. As a result, we have with probability approaching 1:
	$$\lim\inf \theta^* \ge_{(d)} \min\{\sup\Theta^{ID}_{-j}(F_0),M\}=_{(e)} \sup\Theta^{ID}_{-j}(F_0)\ge \lim\sup\theta^*-2\epsilon,$$
	where $(d)$ is established show in \eqref{eq: append, giving up A_j equivalence 1.5}, and $(e)$ follows because $M\ge \sup\Theta^{ID}_{-j}(F_0)$. Since $\epsilon$ is arbitrary, we must have $\theta^*\rightarrow_p \sup  \Theta^{ID}_{-j}(F_0)$. The arguments for $\theta_*$ is similar.
\end{proof}

\section{Proofs in Section \ref{sec: Additional Applications}}

\subsection{Proof of Proposition \ref{prop: bounds for intersection bounds model} }
\begin{proof}
	We first prove the minimal deviation required. Note that for any distribution $G$ of $Y_i,Z_i,\eta_i^+,\eta_i^-$, we have 
	\[\begin{split}
	\eta_i^+=\bar{Y}_i-Y_i,\quad \eta_i^-=Y_i-\underline{Y}_i.
	\end{split}\]
	Then $E[\eta_i^+|Z_i=z]=E[\bar{Y_i}|Z_i=z]-E[Y_i]$, where we use the independence assumption $A_1$ to get $E[Y_i|Z_i]=E[Y_i]$. Similarly, we can get the equality for $E[\eta_i^-|Z_i=_z]$. Therefore, 
	{\footnotesize
	\[
	\left[ \sup_{z\in [z_l,z_u]} E_G[\eta_i^-|Z_i=z]\right]_+  +\left[-\inf_{z\in [z_l,z_u]} E_G[\eta_i^+|Z_i=z]\right]_+=\left[ \sup_{z\in [z_l,z_u]} E_F[\underline{Y}_i|Z_i=z]-\inf_{z\in [z_l,z_u]} E_F[\bar{Y}_i|Z_i=z]\right]_+.
	\]}
	
	We prove the result for $\underline{\theta}(F,m)$ and the upper bound follows by symmetric arguments. By considering $Y_i^*\equiv \inf_{z\in[z_l,z_u]} E_F[\bar{Y}_i|Z_i=z]-m$, and let ${\eta}_i^{+*}\equiv  \bar{Y}_i-Y_i^*$, $\eta^{-*}_i\equiv \underline{Y}_i-Y_i^*$, we can show that 
	\[
	\left[-\inf_{z\in [z_l,z_u]} E_G[\eta_i^{+*}|Z_i=z]\right]_+=0,\quad 
	\left[ \sup_{z\in [z_l,z_u]} E_G[\eta_i^{-*}|Z_i=z]\right]_+=m.
	\]
	Therefore the construction above achieves the lower bound in the proposition. 
	
	Then we show that any $\theta'< \sup_{z\in [z_l,z_u]} E_F[\underline{Y}_i|Z_i=z]-m$, we have a violation so that the bounds in Proposition  \ref{prop: bounds for intersection bounds model}  are sharp. By the construction of $\theta'$, we must have for any $z$:
	\[
	E[\eta_i^{-*}|Z_i=z]\ge E_F[\underline{Y}_i|Z_i=z] -\theta' >	E_F[\underline{Y}_i|Z_i=z]- \sup_{z\in [z_l,z_u]} E_F[\underline{Y}_i|Z_i=z]+m.
	\]
	Take $\sup$ over $z$ on both sides of the above display, we have 
	\[
	\sup_{z} E[\eta_i^{-*}|Z_i=z]>m,
	\]
	which violates the amount of deviation that $m_j(G)=m$.

\end{proof}
\subsection{Proof of Proposition \ref{prop: discrete choice model relaxation}}
\begin{proof}
	\cite{christensen2023counterfactual} shows that  $\underline{\Delta}(\bm{u}^s,\bm{P})$ ($\bar{\Delta}(\bm{u}^s,\bm{P})$) is the minimal (maximal) Chi-squared divergence to rationalize $\bm{u}^s$ with (\ref{eq: discrete choice moment condition}). See equation (16) in 	\cite{christensen2023counterfactual}. Our goal is to show that the constraint set in  in (\ref{eq: discrete choice bounds on mean utility}) and (\ref{eq: discrete choice bounds on mean utility, alternative}) are equivalent.

We first claim that, for fixed $\bm{u}^s$, for any $m\in (\underline{\Delta}(\bm{u}^s,\bm{P}),\bar{\Delta}(\bm{u}^s,\bm{P}))$, we can find a $G_m$ distribution such that: (1) $D(G_m||G_0)=m$; (2) Moment condition (\ref{eq: discrete choice moment condition}) holds. This proves that the constraint set in  (\ref{eq: discrete choice bounds on mean utility, alternative})  implies the constraint set in (\ref{eq: discrete choice bounds on mean utility}).

By the definition of minimal and maximal divergence value and the duality theorem in \cite{christensen2023counterfactual}, we can find an $\epsilon>0$, such that: (1).$m\in[\underline{\Delta}(\bm{u}^s,\bm{P})+\epsilon,\bar{\Delta}(\bm{u}^s,\bm{P})-\epsilon]$. In case $\bar{\Delta}(\bm{u}^s,\bm{P})=+\infty$, define $\bar{\Delta}(\bm{u}^s,\bm{P})-\epsilon$ as an arbitrary number larger than $m$; (2).There is a $\underline{G}_\epsilon$such that $D(\underline{G}_\epsilon||G_0)=\underline{\Delta}(\bm{u}^s,\bm{P})+\epsilon$, a $\bar{G}_\epsilon$ such that $D(\bar{G}_\epsilon||G_0)=\bar{\Delta}(\bm{u}^s,\bm{P})-\epsilon$; and (3).$\bm{u}^s,\underline{G}_\epsilon$, $\bar{G}_\epsilon$ satisfy (\ref{eq: discrete choice moment condition}). 

We consider an $\alpha$ mixture of $\underline{G}_\epsilon$ and $\bar{G}_\epsilon$: $G_\alpha= \alpha \underline{G}_\epsilon + (1-\alpha)\bar{G}_\epsilon$, where $\alpha\in [0,1]$. Therefore, with $\alpha$ probability, individual $i$'s random shocks are drawn from $\underline{G}_\epsilon$, with $(1-\alpha)$ probability, individual $i$'s random shocks are drawn from $\bar{G}_\epsilon$. Since for $\underline{G}_\epsilon$ and $\bar{G}_\epsilon$, (\ref{eq: discrete choice moment condition}) is satisfied and expectation is a linear operator, the $\alpha$-mixture also generate the same choice probabilities and (\ref{eq: discrete choice moment condition}) is also satisfied for any $\alpha$. 

Since $D(G_\alpha||G_0)$ is a continuous function of $\alpha$, and $\lim_{\alpha\rightarrow 0}D(G_\alpha||G_0)=\underline{\Delta}(\bm{u}^s,\bm{P})+\epsilon<m$, $\lim_{\alpha\rightarrow 1}D(G_\alpha||G_0)=\bar{\Delta}(\bm{u}^s,\bm{P})-\epsilon>m$, by intermediate value theorem, we can find $\alpha_m\in (0,1)$ such that $D(G_{\alpha_m}||G_0)=m$. This $G_{\alpha_m}$ will also satisfy the (\ref{eq: discrete choice moment condition}) constraint. We therefore shows that for any $\bm{u}^s$ satisfying the constraints in (\ref{eq: discrete choice bounds on mean utility, alternative}), we can find an $G_{\alpha_m}$ that satisfy the constraints in (\ref{eq: discrete choice bounds on mean utility}). 

Conversely, for any $s=(\bm{u}^s,G^s)$ that satisfies constraints in (\ref{eq: discrete choice bounds on mean utility}), the deviation value $m$ must lie between the minimal and maximal Chi-squared deviation metric as in (\ref{eq: discrete choice bounds on mean utility, alternative}) by Theorem 1 in \cite{christensen2023counterfactual}. So the constraint sets in (\ref{eq: discrete choice bounds on mean utility}) and (\ref{eq: discrete choice bounds on mean utility, alternative}) are equivalent.
\end{proof}
\bibliography{testable_implication_reference}
\bibliographystyle{chicago}

\section*{Appendices and Auxiliary Results for Online Publication}

\section{Proofs of Additional Lemmas Used in Main Proofs}\label{sec: additional proofs in online appendix}

	\subsection{Proof of Lemma \ref{lem: characterizing M(G)=f}}
\begin{proof}
	\textbf{Step 1. $M(G^s)=F$ and $s\in A^{TI}\cap A^{EM-NTAT}$ implies \eqref{eq: append, equiv char of M(G^s)=F}.}
	First, we note that for any $s\in A^{TI}\cap A^{EM-NTAT}$, 
	\begin{equation}\label{eq: append, always taker move mass across z}
		\begin{split}
			g^s(y_1,1,1|1)&=_{(1)}g^s(y_1|D_i(1)=1,D_i(0)=1,Z_i(1)=1)Pr_{G^s}(D_i(1)=D_i(0)=1|Z_i=1)\\
			&=_{(2)}g^s(y_1|D_i(1)=1,D_i(0)=1,Z_i(1)=1)Pr_{G^s}(D_i(1)=D_i(0)=1|Z_i=0)\\
			&=_{(3)}g^s(y_1|D_i(1)=1,D_i(0)=1,Z_i(1)=0)Pr_{G^s}(D_i(1)=D_i(0)=1|Z_i=0)\\
			&=_{(4)} g^s(y_1,1,1|0),
		\end{split}
	\end{equation}
	where $(1),(4)$ follows by conditional expectation, $(2)$ follows by $A^{EM-NTAT}$ and $(3)$ follows by $A^{TI}$. Then, we note that the potential outcome equation \eqref{eq: potential outcome} implies 
	\begin{equation}\label{eq: append, P Q expansion}
		\begin{split}
			P( B_1,1)&=Pr_{G^{s}} (Y_i(1)\in B_1, Y_i(0)\in \mathcal{Y},D_i(1)=1,D_i(0)=1|Z_i=1)\\
			&+Pr_{G^{s}} (Y_i(1)\in B_1, Y_i(0)\in \mathcal{Y},D_i(1)=1,D_i(0)=0|Z_i=1),\\
			Q( B_1,1)&=Pr_{G^{s}} (Y_i(1)\in B_1, Y_i(0)\in \mathcal{Y},D_i(1)=1,D_i(0)=1|Z_i=0)\\
			&+Pr_{G^{s}} (Y_i(1)\in B_1, Y_i(0)\in \mathcal{Y},D_i(1)=0,D_i(0)=1|Z_i=0).
		\end{split}
	\end{equation}  
	Taking the density form of \eqref{eq: append, P Q expansion} and use  \eqref{eq: append, always taker move mass across z} to get the first two lines of \eqref{eq: append, equiv char of M(G^s)=F}.
	
	To derive the last two lines of \eqref{eq: append, equiv char of M(G^s)=F}, we proceed similarly by first showing $g^s(y_0,0,0|1)=g^s(y_0,0,0|0)$ using $A^{TI}$ and $A^{EM-NTAT}$, and then taking the density form of the following potential outcome implications:
	\begin{equation}\label{eq: append, P Q expansion 2}
		\begin{split}
			P( B_0,0)&=Pr_{G^{s}} (Y_i(0)\in B_0, Y_i(1)\in \mathcal{Y},D_i(1)=0,D_i(0)=0|Z_i=1)\\
			&+Pr_{G^{s}} (Y_i(0)\in B_0, Y_i(1)\in \mathcal{Y},D_i(1)=0,D_i(0)=1|Z_i=1),\\
			Q( B_0,0)&=Pr_{G^{s}} (Y_i(0)\in B_0, Y_i(1)\in \mathcal{Y},D_i(1)=1,D_i(0)=0|Z_i=0)\\
			&+Pr_{G^{s}} (Y_i(0)\in B_0, Y_i(1)\in \mathcal{Y},D_i(1)=0,D_i(0)=0|Z_i=0).
		\end{split}
	\end{equation}

	\noindent\textbf{Step 2. The converse. } We construct a joint distribution  $G^*$ of $(Y_i(1),Y_i(0),D_i(1),D_i(0),Z_i)$ such that 
	\begin{enumerate}
		\item $Pr_{G^*}(Z_i=z)=Pr_{F}(Z_i=z)$;
		\item The probability mass of compliers satisfies: $Pr_{G^*}(D_i(1)=1, D_{i}(0)=0|Z_i=1) =\int_{\mathcal{Y}} h_1^s(y)dy$, and $Pr_{G^*}(D_i(1)=1, D_{i}(0)=0|Z_i=0) =\int_{\mathcal{Y}} h^s_0(y)dy$;
		\item The probability mass of defiers satisfies: (i).$Pr_{G^*}(D_i(1)=0, D_{i}(0)=1|Z_i=1) =\int_{\mathcal{Y}} q_F(y,1)-p_F(y,1)+h_1^s(y)dy$, which uses the second line of \eqref{eq: append, equiv char of M(G^s)=F}; and (ii).$Pr_{G^*}(D_i(1)=0, D_{i}(0)=1|Z_i=0) =\int_{\mathcal{Y}} p_F(y,0)-q_F(y,0)+h^s_0(y)dy$, which uses the last line of \eqref{eq: append, equiv char of M(G^s)=F}.
		\item The probability mass of always takes satisfies: $Pr_{G^*}(D_i(1)=D_i(0)=1|Z_i=z)= \int_{\mathcal{Y}}  p_F(y,1)-h_1^s(y)dy$, which uses the first line of \eqref{eq: append, equiv char of M(G^s)=F}. This construction satisfies the $A^{EM-NTAT}$ for always takers;
		\item The probability mass of never takes satisfies: $Pr_{G^*}(D_i(1)=D_i(0)=0|Z_i=z)= \int_{\mathcal{Y}} q_F(y,1)-h^s_0(y)dy$, which uses the third line of \eqref{eq: append, equiv char of M(G^s)=F}. This construction satisfies the $A^{EM-NTAT}$ for never takers;
		\item The conditional density of potential outcomes for compliers satisfies for any measurable sets $B_1$ and $B_0$:
		\[
		Pr_{G^*}(Y_i(1)\in B_1, Y_i(0)\in B_0| D_i(1)=D_i(0)=1,Z_i=z) = \frac{\int_{B_1} h_1^s(y_1)dy_1\times \int_{ B_0}h^s_0(y_0)dy_0}{ \int_{\mathcal{Y}}h_1^s(y_1)dy_1 \int_{\mathcal{Y}}h^s_0(y_0)dy_0 }.
		\]
		Note that the conditional density of potential outcomes is independent of the instrument $Z_i$ given the complier type. This construction satisfies the type independence condition for compliers.
		\item The conditional density of potential outcomes for defiers satisfies for any measurable sets $B_1$ and $B_0$:  
		\[
		\begin{split}
			&\quad Pr_{G^*}(Y_i(1)\in B_1, Y_i(0)\in B_0| D_i(1)=1,D_i(0)=0,Z_i=z) \\
			&= \frac{\int_{B_1} (q_F(y,1)-p_F(y,1)+ h_1^s(y_1))dy_1\times \int_{ B_0}(p_F(y,0)-q_F(y,0)+h^s_0(y_0))dy_0}{ \int_{\mathcal{Y}}(q_F(y,1)-p_F(y,1)+ h_1^s(y_1))dy_1\times \int_{\mathcal{Y}}(p_F(y,0)-q_F(y,0)+h^s_0(y_0))dy_0 }.
		\end{split}
		\]
		This construction satisfies the type independence condition for defiers.
		\item The conditional density of potential outcomes for always takers satisfies for any measurable sets $B_1$ and $B_0$:  
		\[
		\begin{split}
			&\quad Pr_{G^*}(Y_i(1)\in B_1, Y_i(0)\in B_0| D_i(1)=1,D_i(0)=0,Z_i=z) \\
			&= \frac{\int_{B_1} (p_F(y_1,1)-h_1^s(y_1))dy_1\times \int_{ B_0}h_a(y_0)dy_0}{ \int_{\mathcal{Y}}(p_F(y_1,1)- h_1^s(y_1))dy_1 },
		\end{split}
		\]
		where $h_a(y)$ is any distribution density on $\mathcal{Y}$.  This construction satisfies the type independence condition for always takers.
		\item The conditional density of potential outcomes for never takes satisfies for any 
		measurable sets $B_1$ and $B_0$:  
		\[
		\begin{split}
			&\quad Pr_{G^*}(Y_i(1)\in B_1, Y_i(0)\in B_0| D_i(1)=0,D_i(0)=1,Z_i=z) \\
			&= \frac{\int_{B_1} h_n(y_1)dy_1\times \int_{ B_0}(q_F(y,0)-h^s_0(y_0))dy_0}{  \int_{\mathcal{Y}}(q_F(y,0)-h^s_0(y_0))dy_0 },
		\end{split}
		\]
		where $h_n(y)$ is any distribution density on $\mathcal{Y}$. This construction satisfies the type independence condition for defiers.
	\end{enumerate}
	
	The above construction construct a proper distribution $G^*$ and $G^*\in A^{TI}\cap A^{NT-AT}$. 
\end{proof}
\subsection{Proof of Lemma \ref{lem: characterizing m^df(G^s)=m}}
\begin{proof}
	Note that by Lemma \ref{lem: characterizing M(G)=f}, conditional on $Z_i=1$, the density of $y_0$ for defiers is $q_F(y,1)-p_F(y,1)+h_1^s(y)$, and conditional on $Z_i=0$, the density of $y_1$ for defiers is $p_F(y,0)-q_F(y,0)+h_0^s(y)$. Integrating over $y$ we get the conditional probability of defiers, and then multiply the integrals by the corresponding instrument probabilities to get the unconditional probability of defiers.
\end{proof}

\subsection{Proof of Lemma \ref{lem: convergence of integratino on level set}}
\begin{proof}
	By triangular and Jensen inequalities, we have 
	\[\begin{split}
		&\quad \left|\int_{\{y:f_n(y)\ge0\} } f_n(y)dy-\int_{\{y:f_0(y)\ge0\} } f_0(y)dy\right|\\
		&\le  \underbrace{\int_{\{y:f_n(y)\ge0\} } |f_n(y)-f_0(y)|dy}_{\text{Part A}}+\underbrace{\int_{\{y:f_n(y)\ge0,f_0(y)<0\} } |f_0(y)|dy}_{\text{Part B}}+\underbrace{\int_{\{y:f_n(y)<0,f_0(y)\ge0\} } |f_0(y)|dy}_{\text{Part C}}.\\
	\end{split}\]
	
	It is easy to see Part A converges to zero because $f_n\rightarrow f_0$ in $L_1$ norm. 
	
	Part B also converges to zero, because on the region $\{y:f_{n}(y)\ge0,f_0(y)<0\}$, we must have $|f_0(y)|\le |f_n(y)-f_0(y)|$. Therefore
	\[
	\int_{\{y:f_{n}(y)\ge0,f_0(y)<0\} } |f_0(y)|dy \le \int_{\{y:f_{n}(y)\ge0,f_0(y)<0\} } |f_n(y)-f_0(y)|dy\rightarrow 0.
	\]
	The same argument follows for Part C because on $\{y:f_n(y)<0,f_0(y)\ge0\}$, we have $|f_0(y)|\le |f_0(y)-f_n(y)|$. 	
\end{proof}

\section{Monotone IV Model and Choice of $m_j$} \label{sec: monotone IV + choice of m_j}
We now present an application to the monotone instrument model \citep{Mansi2000MonotoneIV}, which illustrates the multiplicity of the choices of deviation metric $m_j$ and the consequences. The econometrician observes the outcome variable $Y_i$, the binary treatment variable $D_i$ and a continuous instrument variable $Z_i$. The outcome variable is generated by the potential outcome model: $Y_i=D_iY_i(1)+(1-D_i)Y_i(0)$. We also assume the exclusion restriction of $Z_i$. Let $F$ be the distribution of $(Y_i,D_i,Z_i)$, and let $G$ denote the distribution of $(Y_i(1),Y_i(0),D_i,Z_i)$. To further simplify the discussion, we assume the instrument is bounded between $z_l$ and $z_u$. Our parameter of interest is the conditional treated outcome at instrumental value $Z_i=z_0$: $\theta(G;z_0)=E_{G}[Y_i(1)|Z_i=z_0]$. We further assume $\theta(G;z)$ as a function of $z$ is continuous.

\begin{figure}
	\centering
	\includegraphics[width=0.9\linewidth]{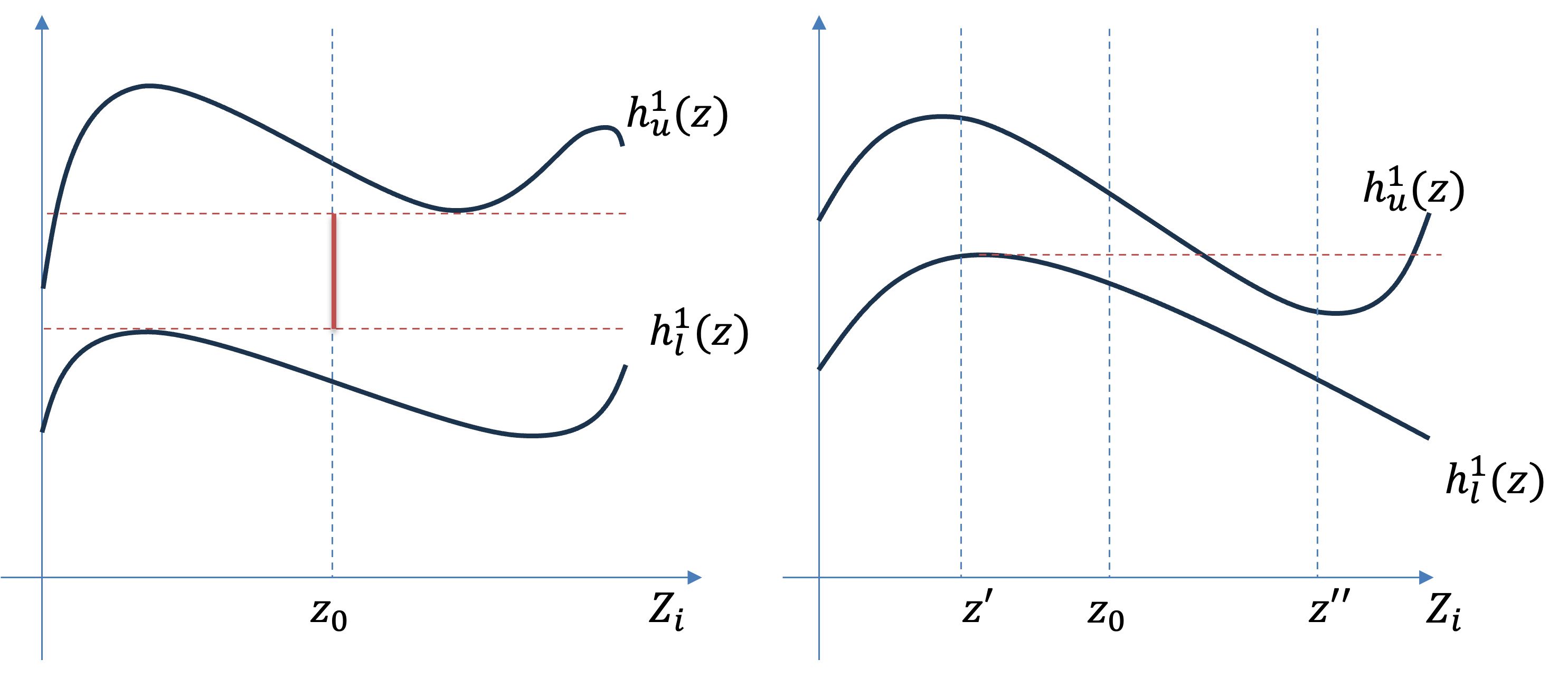}
	\caption{Two Examples.}
	\label{fig:picture-1}
\end{figure}

\cite{Mansi2000MonotoneIV} assume that the  potential outcome is monotonically increasing in $Z_i$ (MIV):
\begin{equation}\label{eq: append, MIV}
	E[Y_i(d)|Z_i=z_1]\ge E[Y_i(d)|Z_i=z_2],\quad \forall z_1\ge z_2,\quad \text{and } d\in \{0,1\}.
\end{equation}
If the econometrician further knows that $Y_i$ is in a bounded interval $[y_l,y_u]$, we can follow \cite{hsu2019testing} to derive  a sharp testable implication.
\begin{lem}\citep{hsu2019testing}\label{lem: HSU 2019}
	Let \[
	\begin{split}
		h^1_u(z)= Pr_F(D_i=0|Z_i=z)y_u+Pr_F(D_i=1|Z_i=z)E[Y_i|D_i=1,Z=z],\\ h^1_l(z)=Pr_F(D_i=0|Z_i=z)y_l+Pr_F(D_i=1|Z_i=z)E[Y_i|D_i=1,Z=z],\\
		h^0_u(z)= Pr_F(D_i=1|Z_i=z)y_u+Pr_F(D_i=0|Z_i=z)E[Y_i|D_i=0,Z=z],\\
		h^0_l(z)=Pr_F(D_i=1|Z_i=z)y_l+Pr_F(D_i=0|Z_i=z)E[Y_i|D_i=1,Z=z].
	\end{split}\]
	 The sharp testable implication of the MIV assumption is 
	\begin{equation}
		h^1_u(z_1)\ge h^1_l(z_2), \quad \text{and}\quad 	h^0_u(z_1)\ge h^0_l(z_2) \quad \forall z_1\ge z_2.
	\end{equation}
\end{lem}
We show two graphical examples in Figure \ref{fig:picture-1} to illustrate the testable implications. The left panel shows an example where the testable implication is not violated, and the identified set of the $E[Y_i(1)|Z_i=z_0]$ is shown as the solid-red-vertial interval.\footnote{See Theorem 1 in \cite{manski2019econometrics}.} The right panel shows an example where the testable implication is violated, since $h_u^1(z'')<h_l^1(z')$. If the right panel is what we observed as a data-identified $h^1_u$ and $h^1_l$, then we can refute the MIV assumption, and the identifid set of $E[Y_i(1)|Z_i=z_0]$ is an empty set. 

We now consider a measure of the deviation from the MIV instrument assumption. Since we only focus on the $E[Y_i(1)|Z_i=z_0]$, we only measure the deviation from the monotonicity of  $E[Y_i(1)|Z_i]$ as illustration. First note that, for any $G^s$ to be consistent with the realized data distribution $(F=M^s(G^s))$, we must have the $G^s$-implied 
\begin{equation}\label{eq: append,  bounds for h_G for MIV}
\begin{split}
	h_{G^s}^1(z)&\equiv E_{G^s}[Y_i(1)|Z_i=z]\\&=Pr_F(D_i=0|Z_i=z)E_G[Y_i(1)|D_i=0,Z_i=z]+Pr_F(D_i=1|Z_i=z)E[Y_i|D_i=1,Z=z],
\end{split}
\end{equation}
which must lie between the $h_l^1(z)$ and $h_u^1(z)$ functions for all $z$ values. We want a measure of the deviation from the assumption \eqref{eq: append, MIV}. 

For any function $h(z)$, we can construct the \textit{left monotone increasing function}, denoted by $h^{left,\uparrow}(z;h)\equiv \sup_{z'\le z} h(z)$. By construction, $h^{left,\uparrow}(z;h)$ is weakly increasing,  and $h(z)\le h^{left,\uparrow}(z;h)$. Moreover, if $h(z)$ is weakly increasing, then $h(z)\equiv h^{left,\uparrow}(z;h)$. Then we can define the deviation metric 
\begin{equation}
	m^{left}_{MIV}(s)= \int_{z_l}^{z_u}  [h^{left,\uparrow}(z;h_{G^s}^1)-h_{G^s}^1(z)] dz.
\end{equation}
Since $h(z)\le h^{left,\uparrow}(z;h)$ holds for all $z$, $m^{left}_{MIV}(s)$ is a mapping to $\mathbb{R}_+$, and equals zero if and only if $h(z)$ is weakly increasing. 

We now show that the deviation metric $m^{left}_MIV$ implies a closed form expressions of $\bar{\theta}(F,m)$ and $\underline{\theta}(F,m)$.  We have the following proposition. 
\begin{prop}\label{prop: MIV deviation m}
	For any $F$ distribution, let $h_l^1(z)$ and $h_u^1(z)$ be defined in Lemma \ref{lem: HSU 2019}. Let $\tilde{h}(z)= \min\{h^{left,\uparrow}(z;h_l^1),h_u^1(z)\}$. Then for any $G^s$ such that $F=M^s(G^s)$, we must have:
	\begin{enumerate}
		\item $m^{left}_{MIV}(G)\ge m^{left}_{MIV}(\tilde{h}(z))\equiv m_{MIV}^{left,min}(F)$, which is the minimal amount of deviation under $m^{left}_{MIV}$; 
		\item For any $\Delta m>0$, $\underline{\theta}(F,m_{MIV}^{left,min}(F)+\Delta m)=h_l^1(z)$ for all $\Delta m> 0$.
		\item 	For any $t$, define $h^*(z;t)= \tilde{h}(z)\mathbbm{1}(z<z_0)+ \max\{t, \tilde{h}(z)\} \mathbbm{1}(z\ge z_0)$, then  
		\[
		\bar{\theta}(F,m_{MIV}^{left,min}(F)+\Delta m)=\max\{t: m_{MIV}^{left}(h^*(z;t))<m_{MIV}^{left,min}(F)+\Delta m\}.
		\]
	\end{enumerate}		
\end{prop}

\begin{figure}
	\centering
	\includegraphics[width=0.9\linewidth]{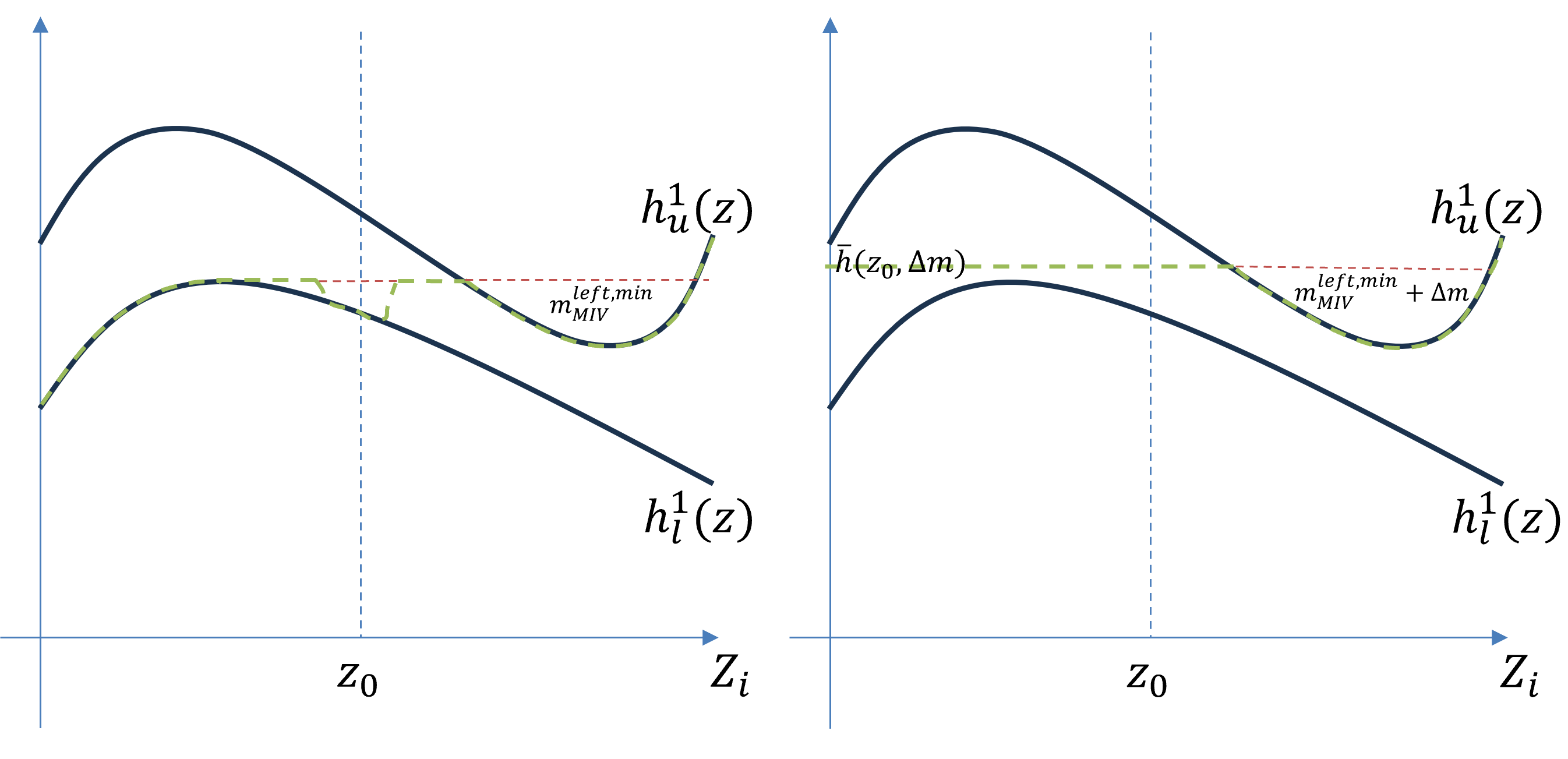}
	\caption{The bounds of $\underline{\theta}(F,m)$ (left panel) and $\bar{\theta}(F,m)$ (right panel) under $m_{MIV}^{left}$.}
	\label{fig:lowerupperboundmiv}
\end{figure}

We first provide some intuitions to Proposition \ref{prop: MIV deviation m}. Let consider $\Delta m$ to be small. We construct the $h_{G^s}$ function that achieves the upper and lower bound in Proposition \ref{prop: MIV deviation m}.  In Figure \ref{fig:lowerupperboundmiv}, the green dashed line in left panel denotes the $h_{G^s}(z)$ function that achieves the lower bound $\bar{\theta}(F,m_{MIV}^{left,min}(F)+\Delta m)$, and the green dashed line in the right panel denotes the $h_{G^s}(z)$ function that achieves the upper bound. For the left panel, we can construct a continuous function which coincide with $h_l^1(z)$ on a neighborhood of $z_0$, and the area between the red-dashed line and the green-dashed line is $\Delta m$.  On the right panel, the region between the red-dashed and green-dashed line is the amount of deviation that we consider: $m_{MIV}^{left,min}(F)+\Delta m$.
When we increase the value of $\Delta m$, to achieve the lower bound $\underline{\theta}(F,m)$ on the left panel, we keep increasing overlapping region of the green dashed line with the $h_l^1(z)$ function; on the right panel, we increase the value of $\bar{h}(z_0,\Delta m)$ until we hit the upper bound $h_u^1(z_0)$. 

There are two things we want to get from this example. First, in many cases, analytical expressions of $\bar{\theta}(F,m)$ and $\underline{\theta}(F,m)$ are available even if the underlying object to maximize over is infinite dimensional. This is manifested in Proposition \ref{prop: MIV deviation m}. Second the choice of the deviation measurement matters. To see this, lets consider a measure of non-monotonicity using the \textit{right monotone increasing function}, denoted by $h^{right,\uparrow}(z;h)\equiv \inf_{z'\ge z} h(z)$. Then, another metric to measure the deviation from assuming monotone $h_G(z)$ is $m_{MIV}^{right}(G)\equiv \int_{z_0}^{z_u} h(z)-h^{right,\uparrow}(z;h) dz$. We can get a different version of Proposition \ref{prop: MIV deviation m}. In particular, the minimal deviation to rationalize $F$ will change to $\int_{z_0}^{z_u} h_u^1(z)-h^{right,\uparrow}(z;h_u^1) dz$, which will lead to different $\bar{\theta}(F,m)$ and $\underline{\theta}(F,m)$.

Depending on the applications,  there may (as in the $LATE$ example) or may not (as in the MIV example) exists a natural way to define the deviation from the baseline assumption $A_j$. In case the choice of deviation metric may result in different  bounds, we can make the interval $[\theta_*,\theta^*]$ robust to the choice of $m_j$ by simply taking the union of all deviation metrics under consideration. 

\subsection{Proof of Proposition \ref{prop: MIV deviation m}}
\begin{proof}
	We first show the minimal deviation required to rationalize the model.  Denote $\mathcal{Z}\equiv \{z: h_u^1(z)\le h^{left,\uparrow}(z;h_l^1)\}$ as the region where the increasing property of $h_u^1$ is violated, which is the overlap region under $m^{left,min}_{MIV}$ in Figure \ref{fig:lowerupperboundmiv}. Then, 
	\[
	m_{MIV}(\tilde{h})= \int_{\mathcal{Z}} h^{left,\uparrow}(z;h_l^1)-h_u^1(z) dz.
	\]

	Now, consider any $h_{G^s}(z)$: By \eqref{eq: append,  bounds for h_G for MIV}, if $F=M^s(G^s)$, we must have $h_{G^s}(z)\ge h_l^1(z)$. Therefore, $h^{left,\uparrow}(z;h_{G^s})\ge h^{left,\uparrow}(z;h_l^1)$ must also hold by the definition of $h^{left,\uparrow}$.  Then 
	\[
	\begin{split}
		m^{left}_{MIV}(G^s)&= \int_{z_l}^{z_u} (h^{left,\uparrow}(z;h_{G^s})-h_{G^s}(z))dz\\
		&\ge \int_{\mathcal{Z}} (h^{left,\uparrow}(z;h_{G^s})-h_{G^s}(z))dz\\
		&\ge_{(i)} \int_{\mathcal{Z}} (h^{left,\uparrow}(z;h_l^1)-h_{G^s}(z))dz\\
		&\ge_{(ii)} \int_{\mathcal{Z}} (h^{left,\uparrow}(z;h_l^1)-h_u^1(z))dz=m_{MIV}(\tilde{h}),	
	\end{split}
	\]
	where $(i)$ follows by  $h^{left,\uparrow}(z;h_{G^s})\ge h^{left,\uparrow}(z;h_l^1)$, and $(ii)$ follows because $h_{G^s}(z)\le h_u^1(z)$. This completes the proof of the first statement of Proposition \ref{prop: MIV deviation m}.
	
	We then prove the lower bound for $E[Y_i(1)|Z_i=z_0]$. Consider any $\Delta m>0$, such that $m_{MIV}^{left,min}(F)+\Delta m$ is smaller than the upper bound of possible deviation. For a small $\epsilon>0$ such that $m_{MIV}^{min}(F)+\Delta m+\epsilon$ is also smaller than the upper bound of possible deviation, we construct a neighborhood $B_{z_0}$ of $z_0$ such that $\int_{B_{z_0}} \tilde{h}(z)-h_l^1(z)dz= \Delta m+\epsilon$. Define $\underline{h}^{dis}(z)= \tilde{h}(z)\mathbbm{1}(z\notin B_{z_0})+ h_l^1(z)\mathbbm{1}(z\in B_{z_0})$. Then $\underline{h}^{dis}(z)$ achieves the lower bound $h_l^1(z_0)$ and $m_{MIV}^{left}(h_l^1)=m_{MIV}^{left,min}(F)+\Delta m+\epsilon$. The $\underline{h}^{dis}(z)$ is not a continuous function, but we can consider any continuous function $h_a(z)$ such that $h_a(z_0)=h_l^1(z_0)$ and $m_{MIV}^{left}(h_a)<m$. Let $\underline{h}^{con}(z;\alpha)=\alpha\underline{h}^{con}(z;\alpha)+(1-\alpha)h_a(z)$, then for any $\alpha\in(0,1)$: (1)$\underline{h}^{con}(z;\alpha)$ is continuous; (2)$\underline{h}^{con}(z_0;\alpha)=h_l^1(z_0)$; (3) By mean value theorem, there exists an $\alpha^*\in(0,1)$ such that $m_{MIV}^{left}(\underline{h}^{con}(\cdot;\alpha^*)=m_{MIV}^{left,min}(F)+\Delta m$. In view of Figure \ref{fig:lowerupperboundmiv}, this $m_{MIV}^{left}(\underline{h}^{con}(\cdot;\alpha^*)$ is the green-dashed line in the left panel. This completes the proof of the second statement.

	To prove the upper bound, for each $t>0$, we first construct a function  $h^*(z;t)$ that is discountinuous at $z_0$ that achieves the bound. The continuously-smoothed approximation can be constructed as the construction of $\underline{h}^{con}(z;\alpha)$ above.  We will show that this $h^*(z;t)$ achieves the maximal functional value at $z_0$ 
	given the amount of deviations.

	
	We first note that if $m_{MIV}^{left}(h^*(z;t))= m^*$, then for any function $h^\#(z)$ such that $h^\#(z_0)>t$, we must have  $m_{MIV}^{left}(h^\#)> m^*$. To see this, we can show that 
	\[
	\begin{split}
		m_{MIV}^{left}(h^\#)&=\int_{z_l}^{z_0} h^{left,\uparrow}(z;h^\#)-h^\#(z) dz +\int_{z_0}^{z_u} h^{left,\uparrow}(z;h^\#)-h^\#(z) dz\\
		&\ge \int_{z_l}^{z_0} h^{left,\uparrow}(z;h_l^1)-h_l^1(z) dz +\int_{z_0}^{z_u} h^{left,\uparrow}(z;h^\#)-h^\#(z) dz\\
	\end{split}
	\]
	where the inequality holds because $h^\#$ must achieve the minimal deviation on the $[z_l,z_0)$ region, which is the integration value $\int_{z_l}^{z_0} h^{left,\uparrow}(z;h_l^1)-h_l^1(z) dz$.
	
	Next, we consider the set $\mathcal{Z}'=\{z\ge z_0: h_u^1(z)< \max\{t,\tilde{h}(z)\} \}$. Then 
	\[
	\begin{split}
		\int_{z_0}^{z_u} h^{left,\uparrow}(z;h^\#)-h^\#(z) dz\ge \int_{\mathcal{Z}'} h^{left,\uparrow}(z;h^\#)-h^\#(z) dz\\
		>_{(iii)} \int_{\mathcal{Z}'} t-h^\#(z) dz\ge_{(iv)} \int_{\mathcal{Z}'} \max\{t,\tilde{h}(z)\}-h^\#(z) dz\\
		\ge_{(v)}  \int_{\mathcal{Z}'} \max\{t,\tilde{h}(z)\}-\tilde{h}(z) dz,
	\end{split}
	\]	
	where $(iii)$ follows by $ h^{left,\uparrow}(z;h^\#)\ge h^\#(z_0)>t$ on $\mathcal{Z}'$, $(iv)$ follows by the definition of $\mathcal{Z}'$, and $(v)$ follows because $\tilde{h}(z)=h_u^1(z)$ on $\mathcal{Z}'$.
	
	Therefore, when computing the $\bar{\theta}(F,m)$, we only need to consider the $h^*(z;t)$ functions. Also note that $min_{MIV}^{left}(h^*(z;t))$ is a weakly continuously increasing function of $t$, so there must exists a $t_0$ where $min_{MIV}^{left}(h^*(z;t_0))=m_{MIV}^{left,min}(F)+\Delta m$. Therefore the maximum can be achieved.
	\end{proof}
\end{document}